\documentclass[a4paper,11pt]{article}%
\usepackage{amssymb}
\usepackage{graphicx}
\usepackage{amsbsy}
\usepackage{amsmath}
\usepackage{cite}
\usepackage{slashed}
\usepackage{amsfonts}
\usepackage{hyperref}%
\hypersetup{
colorlinks   = true,
linkcolor    = blue,
citecolor    = blue,
}

\begin{document}
\thispagestyle{empty} \setcounter{page}{0} \begin{flushright}
June 2020\\
\end{flushright}

\vskip          4.1 true cm

\begin{center}
{\huge Baryon and lepton number intricacies in axion models}\\[1.9cm]

\textsc{J\'er\'emie Quevillon}$^{1}$\textsc{\ and Christopher Smith}$^{2}%
$\vspace{0.5cm}\\[9pt]\smallskip{\small \textsl{\textit{Laboratoire de
Physique Subatomique et de Cosmologie, }}}\linebreak%
{\small \textsl{\textit{Universit\'{e} Grenoble-Alpes, CNRS/IN2P3, Grenoble
INP, 38000 Grenoble, France}.}} \\[1.9cm]\textbf{Abstract}\smallskip
\end{center}

\begin{quote}
\noindent Because the Peccei-Quinn (PQ) symmetry has to be anomalous to solve the strong
CP puzzle, some colored and chiral fermions have to transform non-trivially
under this symmetry. But when the SM fermions are charged, as in the PQ or
DFSZ models, this symmetry ends up entangled with the SM global symmetries,
baryon ($\mathcal{B}$) and lepton ($\mathcal{L}$) numbers. This raises several
questions addressed in this paper. First, the compatibility of axion models
with some explicit $\mathcal{B}$ and/or $\mathcal{L}$ violating effects is
analyzed, including those arising from seesaw mechanisms, electroweak
instanton interactions, or explicit $\mathcal{B}$ and $\mathcal{L}$ violating
effective operators. Second, how many of these effects can be simultaneously
present is quantified, along with the consequences for the axion mass and
vacuum alignment if too many of them are introduced. Finally, large classes of
$\mathcal{B}$ and/or $\mathcal{L}$ violating interactions without impact on axion
phenomenology are identified, like for
example the various implementations of the type I and II seesaw mechanisms in
the DFSZ context.

\let\thefootnote\relax\footnotetext{$^{1}\;$jeremie.quevillon@lpsc.in2p3.fr}%
\ \footnotetext{$^{2}\;$chsmith@lpsc.in2p3.fr}
\end{quote}

\newpage

\setcounter{tocdepth}{2}

\section{Introduction}

Even if the simplest axion models introduced more than forty years ago have
been ruled out, axions still remain one of the best solutions for the strong
CP problem of the Standard Model (SM). This problem originates from
the observation that the QCD and the electroweak sectors, by construction
secluded, must somehow conspire to cancel each other's sources of
CP-violation. Indeed, while individually their contributions to the $\theta$
term of QCD are a priori both of $\mathcal{O}(1)$, the yet non-observed
electric dipole moment of the neutron~\cite{Abel:2020gbr} requires their sum
to be tiny, $\theta_{eff}\equiv\theta_{QCD}+\theta_{Yukawa}\lesssim10^{-10}$.

Axions come under many guises, but the basic receipe is always the same:
design a global $U(1)$ symmetry and assign charges to some colored chiral
fermions~\cite{PQ}. This ensures $U(1)$ rotations act on the strong CP phase
since its current is anomalous. This is not sufficient yet to dispose of the
$\theta$ term since fermion masses explicitly break this $U(1)$ symmetry. To
force $\theta_{eff}$ to zero, the trick proceeds in two steps~\cite{PQ}.
First, this $U(1)$ symmetry is spontaneously broken, so that its associated
Goldstone boson, the axion \cite{Weinberg:1977ma,Wilczek:1977pj}, has a direct
coupling to gluons. Second, non-perturbative QCD effects create an effective
potential for the axion field, whose minimum is attained precisely when the
$\theta$ term is rotated away. In the process, the axion acquires a small
QCD-induced mass, typically well below the eV scale
\cite{Bardeen:1977bd,Kim:1986ax}. Both the mass and the couplings of the QCD
axion are thus controlled by a single scale: the one of spontaneous symmetry
breaking, usually dubbed $f_{a}$.

To solve the strong CP puzzle, the axion needs to be coupled to colored
fermions, and this gives rise to two broad classes of models. Those of the
KSVZ type~\cite{KSVZ} introduce new very heavy fermions, vector-like for the
SM gauge interactions, while those of the PQ~\cite{PQ} and DFSZ~\cite{DFSZ}
types make use of the SM chiral quarks. In that latter case, the axion must
arise from the very same Higgs bosons that give the quarks their masses, and
thus only emerges after the electroweak symmetry is broken. In a previous
study~\cite{Quevillon:2019zrd}, we have described that, for this class of
models, the fermion charges are necessarily ambiguous, because of the presence
of the accidental $U(1)$ symmetries of the SM, corresponding to the conserved
baryon ($\mathcal{B}$) and lepton ($\mathcal{L}$) numbers. Though this
ambiguity was found to have no impact on the low-energy phenomenology, it
raises several questions that we want to address in the present paper. Specifically,

\begin{itemize}
\item Since the ambiguities arise from the SM accidental symmetries, the main
question is to study what happens in the presence of explicit $\mathcal{B}$
and/or $\mathcal{L}$ breaking terms. There is some conflicting conclusions
regarding the capabilities of DFSZ models to accommodate for such violations.
We will see that some limited violation is possible, characterise it, and
study the consequences when this limit is overstepped.

\item A second question is to which extend is it possible to fix the
ambiguities, or said differently, is there naturally some $\mathcal{B}$ and/or
$\mathcal{L}$ components embedded in the axion $U(1)$ symmetry. Of course,
those components are projected out when the symmetry is spontaneously broken,
but finding the optimal representation for the $U(1)$ symmetry could simplify
the form of the axion effective Lagrangian. We will see that in most cases,
neutrino masses and electroweak instanton effects hold the key to identify the
$U(1)$ symmetry unambiguously.

\item Finally, since these ambiguities have no phenomenological consequence,
it is worth to inversigate whether it can be used to relate seemingly
different models. We will see that the fermion charges for all PQ and
DFSZ-like models based on the same Yukawa couplings, whether with a seesaw
mechanism of type I, II, or with some (limited) $\mathcal{B}$ violation, are
actually equivalent. Thus, despite their very different appearance in terms of
effective interactions, those models cannot be distinguished at low energy.
\end{itemize}

The paper is organized as follow. To set the stage, we start in the next
section by presenting the PQ axion model and the DFSZ axion model. Then in
Section~\ref{sec3} we study the compatibility of these models with lepton
number violation, by introducing various mechanisms to generate neutrino
masses. In Section~\ref{sec4} we investigate the impact of baryon number
violation on axion models and explore what would happen if further explicit
$\mathcal{B}$ and/or $\mathcal{L}$ violating interactions were introduced in
the theory. Finally, our results are summarized in Section~\ref{Ccl}.

\section{Fermion charge ambiguities in axion models}

In this section, the simplest axion models are briefly reviewed.\ We focus on
the precise identification of the global and local $U(1)$ symmetries at play,
and their breaking pattern. In this way, it will be immediately obvious that
when the scalars giving masses to the SM fermions are charged under the PQ
symmetry, there remains an ambiguity in the PQ charges of the fermions, and
that this ambiguity is related to the invariance of the Yukawa couplings under
$\mathcal{B}$ and $\mathcal{L}$. In the next sections, this freedom will play
a central role, as it will be used to accommodate the possibility of
$\mathcal{B}$ and/or $\mathcal{L}$ violation in axion models.

\subsection{Axion in the PQ model}

The starting point is a two Higgs doublets with the scalar potential%
\begin{equation}
V_{\text{THDM}}=m_{1}^{2}\Phi_{1}^{\dagger}\Phi_{1}+m_{2}^{2}\Phi_{2}%
^{\dagger}\Phi_{2}+\frac{\lambda_{1}}{2}(\Phi_{1}^{\dagger}\Phi_{1})^{2}%
+\frac{\lambda_{2}}{2}(\Phi_{2}^{\dagger}\Phi_{2})^{2}+\lambda_{3}(\Phi
_{1}^{\dagger}\Phi_{1})(\Phi_{2}^{\dagger}\Phi_{2})+\lambda_{4}(\Phi
_{2}^{\dagger}\Phi_{1})(\Phi_{1}^{\dagger}\Phi_{2})\;.
\end{equation}
Provided a consistent Spontaneous Symmetry Breaking (SSB) occurs, the mass
spectrum is then made of two neutral scalar Higgs bosons $h^{0}$ and $H^{0}$,
a pseudoscalar $A^{0}$, and a pair of charged Higgs boson $H^{\pm}$.

This potential is invariant under the independent rephasing of the Higgs
doublets, corresponding to a global $U(1)_{1}\otimes U(1)_{2}$ symmetry.
Actually, a linear combination of these $U(1)$ charges is nothing but the
gauged hypercharge. Note that this $U(1)_{1}\otimes U(1)_{2}$ symmetry is
truly active at the level of the whole THDM, and in particular, assuming
Yukawa couplings of Type II,%
\begin{equation}
\mathcal{L}_{\text{Yukawa}}=-\bar{u}_{R}\mathbf{Y}_{u}q_{L}\Phi_{1}-\bar
{d}_{R}\mathbf{Y}_{d}q_{L}\Phi_{2}^{\dagger}-\bar{e}_{R}\mathbf{Y}_{e}\ell
_{L}\Phi_{2}^{\dagger}+h.c.\;, \label{YukQuark}%
\end{equation}
it requires that fermions are assigned appropriate $U(1)_{1}\otimes U(1)_{2}$
charges. Beside, these Yukawa couplings are also invariant under the global
baryon and lepton number symmtries, $U(1)_{\mathcal{B}}$ and
$U(1)_{\mathcal{L}}$. Those must be left untouched by the Electroweak Symmetry
Breaking (EWSB). So, all in all, the pattern of symmetry breaking is%
\begin{align}
G_{THDM}  &  =U(1)_{\mathcal{B}}\otimes U(1)_{\mathcal{L}}\otimes
U(1)_{1}\otimes U(1)_{2}\otimes SU(2)_{L}\otimes SU(3)_{C}\nonumber\\
&  =U(1)_{\mathcal{B}}\otimes U(1)_{\mathcal{L}}\otimes U(1)_{X}\otimes
U(1)_{Y}\otimes SU(2)_{L}\otimes SU(3)_{C}\nonumber\\
&  \xrightarrow{\text{EWSB}} U(1)_{\mathcal{B}}\otimes U(1)_{\mathcal{L}%
}\otimes U(1)_{em}\otimes SU(3)_{C}\ . \label{PatternTHDM}%
\end{align}
When the doublets acquire Vacuum Expectation Values (VEVs), $U(1)_{1}\otimes
U(1)_{2}\otimes SU(2)_{L}$ is broken down to $U(1)_{em}$. There are thus two
Goldstone bosons, one is the Would-be Goldstone (WBG) eaten by the $Z^{0}$,
and the other is truly present in the spectrum and is the massless axion.

In the breaking chain, it must be stressed that we wrote $U(1)_{X}$ and not
$U(1)_{PQ}$ for the part of $U(1)_{1}\otimes U(1)_{2}$ not aligned with
$U(1)_{Y}$. Indeed, strictly speaking, the $U(1)_{PQ}$ symmetry is only
defined after the doublets acquire their VEVs, from the orthogonality of the
axion with the WBG of the $Z^{0}$. Further, if we denote the VEVs as
$\langle0|\operatorname{Re}\Phi_{i}|0\rangle=v_{i}$ with $v_{1}^{2}+v_{2}%
^{2}\equiv v^{2}\approx\left(  246\,\text{GeV}\right)  ^{2}$ and $v_{2}%
/v_{1}\equiv x\equiv1/\tan\beta$, both these fields are $v_{i}$-dependent
linear combinations of $\operatorname{Im}\Phi_{1}^{0}$ and $\operatorname{Im}%
\Phi_{2}^{0}$, and consequently, the PQ charges of the doublets are functions
of $v_{i}$. They are only defined once $U(1)_{Y}$ is broken.

Specifically, adopting a polar representation for the pseudoscalar Goldstone
bosons, the Higgs doublets are written in the broken phase as%
\begin{equation}
\Phi_{i}=\frac{1}{\sqrt{2}}\exp(i\eta_{i}/v_{i})\left(
\begin{array}
[c]{c}%
H_{i}^{+}\\
v_{i}+\operatorname{Re}H_{i}%
\end{array}
\right)  \;,\;i=1,2\;. \label{PolPara}%
\end{equation}
The Goldstone bosons associated to the $U(1)_{1}$ and $U(1)_{2}$ symmetries,
$\eta_{1}$ and $\eta_{2}$, are related to the physical Goldstone bosons
$a^{0}$ and $G^{0}$ as%
\begin{equation}
\left(
\begin{array}
[c]{c}%
G^{0}\\
a^{0}%
\end{array}
\right)  =\left(
\begin{array}
[c]{cc}%
\cos\beta & \sin\beta\\
-\sin\beta & \cos\beta
\end{array}
\right)  \left(
\begin{array}
[c]{c}%
\eta_{2}\\
\eta_{1}%
\end{array}
\right)  \ . \label{Rotbeta}%
\end{equation}
Plugging this in Eq.~(\ref{PolPara}), the PQ charge of each doublet can be
read off its phase variation under a shift of the associated Goldstone boson,
$a^{0}\rightarrow a^{0}+v\theta$, and thus%
\begin{equation}
PQ(\Phi_{1})=\frac{v}{v_{1}}\cos\beta=x\ ,\ \ PQ(\Phi_{2})=-\frac{v}{v_{2}%
}\sin\beta=-\frac{1}{x}\ . \label{PQHiggs}%
\end{equation}
Note that the shift $G^{0}\rightarrow G^{0}+v\theta$ reproduces $Y(\Phi
_{1})=Y(\Phi_{2})=1$. It also shows explicitly how misleading any idea of
orthogonality of the $U(1)$ charges could be. We started with $U(1)_{1}\otimes
U(1)_{2}$ under which the pair $(\Phi_{1},\Phi_{2})$ has the seemingly
orthogonal charge assignment $(v/v_{1},0)\otimes(0,v/v_{2})$.\ But once
$U(1)_{1}\otimes U(1)_{2}$ is broken and the associated Goldstone bosons
compelled to be orthogonal, we end up with the $U(1)_{Y}\otimes U(1)_{PQ}$
charge $(1,1)\otimes(x,-1/x)$ for the pair $(\Phi_{1},\Phi_{2})$.

Once these charges are fixed, those of the fermions can be derived by
requiring the Yukawa Lagrangian to be invariant under $U(1)_{PQ}$. Since those
couplings are also necessarily invariant under $\mathcal{B}$ and $\mathcal{L}%
$, these charges are defined only up to a two-parameter
ambiguity~\cite{Quevillon:2019zrd}, which we denote $\alpha$ and $\beta$:%
\begin{equation}
PQ(q_{L},u_{R},d_{R},\ell_{L},e_{R})=(\alpha,\alpha+x,\alpha+\frac{1}{x}%
,\beta,\beta+\frac{1}{x})\ \ . \label{PQferm}%
\end{equation}
At this stage, there is no way to fix $\alpha$ and $\beta$, essentially
because neither $\mathcal{B}$ nor $\mathcal{L}$ have associated dynamical
fields. Further, as discussed for the pair $(\Phi_{1},\Phi_{2})$, there is no
viable concept of orthogonality for the $U(1)$ charges in the fermion sector
either. Actually, it should be remarked that
\begin{subequations}
\begin{align}
\mathcal{B}(q_{L},u_{R},d_{R},\ell_{L},e_{R})  &  =(1/3,1/3,1/3,0,0)\ ,\\
\mathcal{L}(q_{L},u_{R},d_{R},\ell_{L},e_{R})  &  =(0,0,0,1,1)\ ,\\
Y(q_{L},u_{R},d_{R},\ell_{L},e_{R})  &  =(1/3,4/3,-2/3,-1,-2)\ , \label{Y}%
\end{align}
are not orthogonal among themselves to begin with, so there is no reason to
expect the PQ charge to be any different.

The freedom in the PQ charges of the SM fermions has no observable
consequence. The simplest way to see that is to adopt the usual linear
parametrization for the THDM. Since the ambiguity in the fermion PQ charges
appears nowhere in the Lagrangian, all the Feynman rules are independent of
$\alpha$ and $\beta$, and so are the physical observables. Using the polar
representation of Eq.~(\ref{PolPara}), the situation is a bit more involved.
Though initially, the Lagrangian is again independent of $\alpha$ and $\beta$,
and so are all the Feynman rules, it is customary to perform a
reparametrization of the fermion fields to remove the axion field from the
Yukawa couplings. In full generality, this reparametrization is $\alpha$ and
$\beta$ dependent because the fermion rephasings are tuned by their PQ
charges,
\end{subequations}
\begin{equation}
\psi\rightarrow\psi\exp(iPQ(\psi)a^{0}/v)\ ,\ \psi=q_{L},u_{R},d_{R},\ell
_{L},e_{R}\ . \label{freparam}%
\end{equation}
In this way, a dependence on $\alpha$ and $\beta$ is spuriously introduced in
the Lagrangian, first because the non-invariance of the fermion kinetic terms
generates the couplings
\begin{equation}
\delta\mathcal{L}_{\text{\textrm{Der}}}=-\frac{\partial_{\mu}a^{0}}{v}%
\sum_{\psi=q_{L},u_{R},d_{R},\ell_{L},e_{R}}PQ(\psi)\bar{\psi}\gamma^{\mu}%
\psi\ ,
\end{equation}
and second, because the non-invariance of the fermionic path integral measure
generates the anomalous interactions%
\begin{equation}
\delta\mathcal{L}_{\text{\textrm{Jac}}}=\frac{a^{0}}{16\pi^{2}}\left\{
\mathcal{N}_{C}g_{s}^{2}G_{\mu\nu}^{a}\tilde{G}^{a,\mu\nu}+\mathcal{N}%
_{L}g^{2}W_{\mu\nu}^{i}\tilde{W}^{i,\mu\nu}+\mathcal{N}_{Y}g^{\prime2}%
B_{\mu\nu}\tilde{B}^{\mu\nu}\right\}  \label{dJPQTHDM}%
\end{equation}
with
\begin{subequations}
\label{JPQTHDM}%
\begin{align}
\mathcal{N}_{C}  &  =\sum_{\ \ \ \psi=q_{L}^{\dagger},u_{R},d_{R}\ \ \ }%
d_{L}(\psi)C_{C}(\psi)PQ(\psi)=\frac{1}{2}\left(  x+\frac{1}{x}\right)  \ ,\\
\mathcal{N}_{L}  &  =\sum_{\ \ \ \ \ \psi=q_{L}^{\dagger},\ell_{L}^{\dagger
}\ \ \ \ \ }d_{C}(\psi)C_{L}(\psi)PQ(\psi)=-\frac{1}{2}(3\alpha
+\beta)\ ,\\
\mathcal{N}_{Y}  &  =\sum_{\psi=q_{L}^{\dagger},u_{R},d_{R},\ell_{L}^{\dagger
},e_{R}}d_{L}(\psi)d_{C}(\psi)C_{Y}(\psi)PQ(\psi)=\frac{1}{2}\left(
3\alpha+\beta\right)  +\frac{4}{3}\left(  x+\frac{1}{x}\right)\ ,
\end{align}
where $d_{C,L}(\psi)$, $C_{C,L}(\psi)$ are the $SU(3)_{C}$ and $SU(2)_{L}$
dimensions and quadratic Casimir invariant of the representation carried by
the field $\psi$, respectively, and by extension, $C_{Y}(\psi)=Y(\psi)^{2}/4$
with the hypercharges given in Eq.~(\ref{Y}).

Yet, even if both $\delta\mathcal{L}_{\text{\textrm{Der}}}$ and $\delta
\mathcal{L}_{\text{\textrm{Jac}}}$ depend on $\alpha$ and $\beta$, these
parameters cancel out systematically in all physical observables, as shown
explicitly in Ref.~\cite{Quevillon:2019zrd}. Nevertheless, some theoretical
quantities inevitably depend on $\alpha$ and $\beta$. Besides the above
interactions, another particular example is the divergence of the PQ current
since it is related to the anomalous interaction via $\delta\mathcal{L}%
_{\text{\textrm{Jac}}}=a^{0}\partial_{\mu}J_{PQ}^{\mu}$.

Since the two-photon coupling arises as $\mathcal{N}_{L}+\mathcal{N}%
_{Y}=\mathcal{N}_{em}$ in Eq.~(\ref{dJPQTHDM}), both the QED and QCD terms in
$\partial_{\mu}J_{PQ}^{\mu}$ are independent of $\alpha$ and $\beta$ and
immediately physical, but the electroweak term is always ambiguous. This is of
course expected in view of the $\mathcal{B}$ and $\mathcal{L}$ origins of the
$\alpha$ and $\beta$ parameters. If one remembers that these currents also
have anomalous divergences
\end{subequations}
\begin{equation}
\partial_{\mu}J_{\mathcal{B}}^{\mu}=\partial_{\mu}J_{\mathcal{L}}^{\mu}%
=-\frac{N_{f}}{16\pi^{2}}\left(  \frac{1}{2}g^{2}W_{\mu\nu}^{i}\tilde
{W}^{i,\mu\nu}-\frac{1}{2}g^{\prime2}B_{\mu\nu}\tilde{B}^{\mu\nu}\right)  \ ,
\label{dJBdJL1}%
\end{equation}
one can immediately understand how $\alpha$ and $\beta$ enters in
Eq.~(\ref{JPQTHDM}). Yet, one should not conclude too quickly that $\alpha$
and $\beta$ represent a spurious $\mathcal{B}$ and $\mathcal{L}$ component of
the PQ current and should be set to zero. Indeed, this would entirely remove
the electroweak $W_{\mu\nu}^{i}\tilde{W}^{i,\mu\nu}$ term of $\partial_{\mu
}J_{PQ}^{\mu}$, but there is no reason for an (hypothetical) $\mathcal{B}$ and
$\mathcal{L}$-free PQ current to have no electroweak component. Besides, one
should realize that the final form of $\mathcal{N}_{L}$ reflects the specific
choice made in parametrizing the two-parameter freedom in the fermion PQ
charges. To bring Eq.~(\ref{PQferm}) to a simple form, we made the choice of
fixing $PQ(q_{L})\equiv\alpha$ and $PQ(\ell_{L})\equiv\beta$. So, setting
$\alpha=\beta=0$ would simply removes the left-handed fields from the PQ
currents, but this is hardly natural since the axion is coupled to left-handed
fields, as can be confirmed adopting the usual linear representation for the
THDM scalar fields.

\subsection{Axion in DFSZ model}

When the axion is embedded as one of the pseudoscalar degrees of freedom of
the THDM, its couplings end up tuned by the electroweak VEV and are far too
large given the experimental constraints. The DFSZ axion model~\cite{DFSZ}
circumvents this problem by moving most of the axion field into a new field,
whose dynamics take place at a much higher scale. Specifically, the THDM is
extended by a gauge-singlet complex scalar field $\phi$, with the scalar
potential%
\begin{equation}
{V_{\text{DFSZ}}=V_{\text{THDM}}+V_{\phi}+V_{\phi\text{THDM}}+V_{\phi
\text{PQ}}\ ,\ }\left\{
\begin{array}
[c]{l}%
V_{\phi}=\mu^{2}\phi^{\dagger}\phi+\lambda(\phi^{\dagger}\phi)^{2}\dfrac{{}%
}{{}}\ ,\\
V_{\phi\text{THDM}}=a_{1}\phi^{\dagger}\phi\Phi_{1}^{\dagger}\Phi_{1}%
+a_{2}\phi^{\dagger}\phi\Phi_{2}^{\dagger}\Phi_{2}\ \dfrac{{}}{{}},\\
V_{\phi\text{PQ}}=-\lambda_{12}\phi^{2}\Phi_{1}^{\dagger}\Phi_{2}%
+h.c.\;\dfrac{{}}{{}}.
\end{array}
\right.
\end{equation}
This potential is invariant under the same $U(1)_{1}\otimes U(1)_{2}$ symmetry
as in the PQ realization of the previous section, provided $\phi$ is charged
under both $U(1)$s. Concerning fermions, the same type II Yukawa couplings as
in Eq.~(\ref{YukQuark}) are allowed, while $\phi$ cannot directly couple to
fermions because of its $U(1)$s charges.

The symmetry-breaking scale $v_{s}$ of the singlet is assumed to be far above
the electroweak scale. To leading order in $v/v_{s}$, $\langle
0|\operatorname{Re}\phi|0\rangle$ breaks $U(1)_{1}\otimes U(1)_{2}\rightarrow
U(1)_{Y}$ and its associated Goldstone boson is the axion. Indeed, at the $v$
scale, the $\lambda_{12}\langle\phi^{2}\rangle\Phi_{1}^{\dagger}\Phi_{2}$ term
ensures the pseudoscalar state of the THDM is massive. In this leading order
approximation, the axion is not coupled to fermions since it is fully embedded
in $\phi$. The interesting physics take place at $\mathcal{O}(v/v_{s})$, where
the $V_{\phi\text{PQ}}$ coupling generates an $\mathcal{O}(v/v_{s})$ mass for
$\operatorname{Im}\phi$ tuned by $\lambda_{12}v_{1}v_{2}$. Neither
$\operatorname{Im}\phi$ nor $\operatorname{Im}\Phi_{1,2}$ remain massless, but
a linear combination of these states does. The axion is thus $a^{0}%
=\mathcal{O}(1)\operatorname{Im}\phi+\mathcal{O}(v/v_{s})\operatorname{Im}%
\Phi_{1,2}$, and since all the couplings to SM particles stem from its
$\operatorname{Im}\Phi_{1,2}$ components, the axion essentially but not
totally decouples. Yet, it is still able to solve the strong CP problem since
this ensures its coupling to $G_{\mu\nu}\tilde{G}^{\mu\nu}$.

To be more quantitative, this picture is easily confirmed adopting a polar
representation for the scalar fieds.\ Plugging Eq.~(\ref{PolPara}) together
with%
\begin{equation}
\phi=\frac{1}{\sqrt{2}}\exp(i\eta_{s}/v_{s})(v_{s}+\sigma_{S})\ ,
\label{PolPhi}%
\end{equation}
in $V_{\text{DFSZ}}$ and setting all fields but $\eta_{1,2,s}$ to zero, only
the $\lambda_{12}\phi^{2}\Phi_{1}^{\dagger}\Phi_{2}$ coupling contributes
since all the other terms involve the hermitian combinations $\Phi
_{i}^{\dagger}\Phi_{i}$ and/or $\phi^{\dagger}\phi$. Restricted to the
pseudoscalar states, the potential collapses to%
\begin{equation}
V_{\text{DFSZ}}(\eta_{1,2,s})=-\frac{1}{2}\lambda_{12}v_{1}v_{2}v_{s}^{2}%
\cos\left(  \frac{\eta_{1}}{v_{1}}-\frac{\eta_{2}}{v_{2}}-\frac{2\eta_{s}%
}{v_{s}}\right)  \ .
\end{equation}
By expanding the cosine function and diagonalizing the quadratic term, the
mass eigenstates are easily found to be%
\begin{equation}
\left(
\begin{array}
[c]{c}%
G^{0}\\
a^{0}\\
\pi^{0}%
\end{array}
\right)  =\left(
\begin{array}
[c]{ccc}%
\sin\beta & \cos\beta & 0\\
\delta_{s}\omega\cos\beta\sin2\beta & -\delta_{s}\omega\sin\beta\sin2\beta &
\omega\\
\omega\cos\beta & -\omega\sin\beta & -\delta_{s}\omega\sin2\beta
\end{array}
\right)  \left(
\begin{array}
[c]{c}%
\eta_{1}\\
\eta_{2}\\
\eta_{s}%
\end{array}
\right)  \ , \label{RotDFSZ}%
\end{equation}
with $\delta_{s}=v/v_{s}$ and $\omega^{-2}=1+\delta_{s}^{2}\sin^{2}2\beta$.
The interest of this form is that we can read off the PQ charges of $\eta_{1}%
$, $\eta_{2}$, and $\eta_{s}$ from their reactions to a shift $a^{0}%
\rightarrow a^{0}+v_{s}\omega^{-1} \theta$, and we find%
\begin{equation}
PQ(\Phi_{1},\Phi_{2},\phi)=(2\cos^{2}\beta\ ,\ -2\sin^{2}\beta\ ,\ 1\ )\ ,
\end{equation}
or, rescaling these charges by $2x/(x^{2}+1)$,
\begin{equation}
PQ(\Phi_{1},\Phi_{2},\phi)=\left(  x\ ,\ -\frac{1}{x}\ ,\ \frac{1}{2}\left(
x+\frac{1}{x}\right)  \right)  \ . \label{PQdfsz}%
\end{equation}
We thus recover the same charges as in the PQ model, Eq.~(\ref{PQHiggs}), so
those of the fermions also stay the same, Eq.~(\ref{PQferm}), including the
$\alpha$ and $\beta$ ambiguities related to baryon and lepton numbers. Note
that the final form of the mixing matrix is compatible with the cosine
potential, in the sense that the massive $\pi^{0}$ state is precisely the
combination of states occurring as argument of the cosine function:%
\begin{equation}
\pi^{0}=\omega\left(  \cos\beta\eta_{1}-\sin\beta\eta_{2}-\delta_{s}\sin
2\beta\eta_{s}\right)  =\omega v\sin\beta\cos\beta\left(  \frac{\eta_{1}%
}{v_{1}}-\frac{\eta_{2}}{v_{2}}-2\frac{\eta_{s}}{v_{s}}\right)  \ .
\end{equation}
The potential $V_{\text{DFSZ}}(\eta_{1,2,s})$ is necessarily flat in the other
two orthogonal directions, corresponding to the two Goldstone bosons (the
$G^{0}$ eaten by the $Z^{0}$, and the $a^{0}$). Finally, remark that if the
$\phi^{2}\Phi_{1}^{\dagger}\Phi_{2}$ coupling is replaced by $\phi\Phi
_{1}^{\dagger}\Phi_{2}$, everything stays the same but for $PQ(\phi)$.\ This
has no phenomenological impact since the axion couplings to SM fields are unchanged.

\section{Axions and lepton number violation\label{sec3}}

Up to now, neutrinos have been kept massless. To account for the very light
neutrino masses in a natural way, the standard approach is to implement a
seesaw mechanism. Generically, these mechanisms assume the observed
left-handed neutrinos have a Majorana mass term, typically via the
dimension-five operator%
\begin{equation}
\mathcal{L}_{seesaw}^{eff}=-\frac{\mathbf{c}}{\Lambda}(\bar{\ell}_{L}^{C}%
\Phi_{i}^{T})(\ell_{L}\Phi_{i})+h.c. \label{leffseesaw}%
\end{equation}
where $\mathbf{c}$ is understood as a matrix in flavor space, and flavor
indices are understood. Neutrino masses are then $\mathbf{m}_{\nu}%
=\mathbf{c}v_{i}^{2}/\Lambda$. The scale $\Lambda$ represents that where
lepton number is broken, either explicitly or spontaneously. Obviously,
neutrinos end up very light when $\Lambda$ is sufficiently high.

Since a generic feature of the seesaw mechanisms is a breaking of
$\mathcal{L}$, the most immediate question is how to accommodate for that in
axion models. This has already been studied quite extensively, but most of the
time in a KSVZ-like setting, where new colored fermions are introduced and SM
fermions need not be charged under the PQ
symmetry{~\cite{Shin:1987xc,Ballesteros:2016xej,Demir:2000fj,Ma:2017vdv}}.
Here, we concentrate on DFSZ-like models, in which $\mathcal{L}$ manifests
itself as an ambiguity in the PQ charges of the SM fermions.\ To study the
consequences, and actually show that axion phenomenology is essentially
unaffected by neutrino masses, we review in this section three realizations.
First, we supplement the PQ and DFSZ model with a seesaw mechanism of type
I~\cite{TypeI}. Then, we consider the $\nu$DFSZ model of
Ref.~\cite{Clarke:2015bea}, where the DFSZ singlet is made responsible for the
breaking of lepton number. Finally, we consider the type II seesaw
mechanism~\cite{TypeII}, realized either \`{a} la PQ or
DFSZ~\cite{Bertolini:1990vz,Bertolini:2014aia}. Other DFSZ-like realizations
are possible, see for example
Refs.{~\cite{Mohapatra:1982tc,He:1988dm,Gu:2016hxh,Ahn:2015pia}, }but those
described here are the simplest. Also, we do not consider the proposal of
Ref.~{\cite{Latosinski:2012qj,Heeck:2019guh}} in which the PQ and
$\mathcal{B}-\mathcal{L}$ currents are identified, with a non-local majoron
gluonic coupling arising through complicated multiloop processes.

\subsection{PQ and DFSZ with a type I seesaw mechanism}

A first strategy to account for neutrino masses is to add to the PQ or DFSZ
model a type I seesaw mechanism. Specifically, we add right-handed neutrinos
$\nu_{R}$ to the model. Since those are singlet under the gauge symmetry, the
only new allowed couplings are%
\begin{equation}
\mathcal{L}_{\nu_{R}}=-\frac{1}{2}\bar{\nu}_{R}^{C}\mathbf{M}_{R}\nu_{R}%
+\bar{\nu}_{R}\mathbf{Y}_{\nu}\ell_{L}\Phi_{i}+h.c.\;. \label{mnur}%
\end{equation}
with $i=1$ or $2$.\ Lepton number no longer emerges as an accidental symmetry
because the Majorana mass term $\mathbf{M}_{R}$ breaks $\mathcal{L}$ by two
units. It is also presumably very large, so integrating out the $\nu_{R}$
fields generates the dimension-five operator in Eq.~(\ref{leffseesaw}) with
$\mathbf{c}\Lambda^{-1}=\mathbf{Y}_{\nu}^{T}\mathbf{M}_{R}^{-1}\mathbf{Y}%
_{\nu}$.

The PQ charge of the right handed neutrinos has to vanish to allow the
presence of the Majorana mass term. Given the PQ charge in Eqs.~(\ref{PQferm})
and~(\ref{PQHiggs}) or~(\ref{PQdfsz}), this implies that $\beta$ must be
non-zero since
\begin{subequations}
\label{betaPQ}%
\begin{align}
\bar{\nu}_{R}\mathbf{Y}_{\nu}\ell_{L}\Phi_{1}  &  :PQ(\nu_{R})=\beta+x=0\ ,\\
\bar{\nu}_{R}\mathbf{Y}_{\nu}\ell_{L}\Phi_{2}  &  :PQ(\nu_{R})=\beta-\frac
{1}{x}=0\ .
\end{align}
These equations must be interpreted in the right way. This is not a choice for
$\beta$. Rather, in the presence of $\mathbf{M}_{R}$, $U(1)_{\mathcal{L}}$ is
removed from the symmetry breaking chain of Eq.~(\ref{PatternTHDM}), and the
corresponding ambiguity is simply not there to start with. In other words, it
would make no sense to set $\beta$ to any other value and discuss the impact
of the PQ breaking induced by $\mathbf{M}_{R}$, since this breaking is
spuriously introduced by an inappropriate choice of PQ charges. Yet,
remarkably, the PQ symmetry does not forbid either the Majorana mass term in
Eq.~(\ref{mnur}) or the effective operator Eq.~(\ref{leffseesaw}), contrary to
the claim made for example in Refs.~\cite{Peinado:2019mrn,Baek:2019wdn}.

The presence of the seesaw mechanism does not significantly alter the axion
phenomenology. This is most clearly seen adopting the linear parametrization
for the scalar fields, since then all the axion couplings to SM fermions are
proportional to their masses. When $\nu_{R}$ have been integrated out, that of the axion to light neutrinos will arise from
Eq.~(\ref{leffseesaw}), and thus be tiny.
In a polar representation for the pseudoscalar fields, first note that with
$\beta$ fixed as in Eq.~(\ref{betaPQ}), the seesaw operator of
Eq.~(\ref{leffseesaw}) becomes invariant under the PQ symmetry. It does not
prevent the fermion reparametrization of Eq.~(\ref{freparam}), which proceeds
exactly as in the absence of neutrino masses. Except that $\beta$ is fixed,
the effective derivative and anomalous interactions stay the same.\ Since we
proved in Ref.~\cite{Quevillon:2019zrd} that $\beta$ cancels out in physical
observables anyway, the phenomenology is unchanged, except for the tiny
kinematical impact of the now finite neutrino masses (for example, the
$a^{0}W^{+}W^{-}$ loop amplitude depends on the mass of the virtual fermions,
including neutrinos).

\subsection{Merging DFSZ with a type I seesaw mechanism}

Instead of adding a Majorana mass term for the right-handed neutrinos, we can
use the singlet field and set%
\end{subequations}
\begin{equation}
\mathcal{L}_{\nu_{R}}=-\frac{1}{2}\bar{\nu}_{R}^{C}\mathbf{Y}_{R}\nu_{R}%
\phi+\bar{\nu}_{R}\mathbf{Y}_{\nu}\ell_{L}\Phi_{i}+h.c.\;.
\label{SingletSeesaw}%
\end{equation}
This model, dubbed the $\nu$DFSZ, was first proposed in
Ref.~\cite{Clarke:2015bea}.

Let us see how this merging of the DFSZ model with a type I seesaw mechanism
can be understood from the point of view of the $U(1)$s. Since $\phi$ cannot
be neutral under $U(1)_{1}\otimes U(1)_{2}$, the right-handed neutrinos do
have charges, and no Majorana mass term is allowed. Basically, what we are
doing is to embed lepton number inside the global symmetries,
$U(1)_{\mathcal{L}}\subset U(1)_{1}\otimes U(1)_{2}$. Since the VEV of $\phi$
breaks both $U(1)_{1}$ and $U(1)_{2}$, it also breaks $U(1)_{\mathcal{L}}$,
and then the Goldstone boson can be viewed as a majoron. Note, though, that
$\Phi_{1}$ and $\Phi_{2}$ as well as quarks are charged under $U(1)_{1}\otimes
U(1)_{2}$, since the assignments are%
\begin{equation}%
\begin{tabular}
[c]{cccccccccc}\hline
$\bar{\nu}_{R}\mathbf{Y}_{\nu}\ell_{L}\Phi_{1}$ & $\phi$ & $\Phi_{1}$ &
$\Phi_{2}$ & $q_{L}$ & $u_{R}$ & $d_{R}$ & $\ell_{L}$ & $e_{R}$ & $\nu_{R}%
$\\\hline
$\ \ U(1)_{1}\ \ $ & $+1/2$ & $1$ & $0$ & $\alpha_{1}$ & $\alpha_{1}+1$ &
$\alpha_{1}$ & $-5/4$ & $-5/4$ & $-1/4$\\
$\ \ U(1)_{2}\ \ $ & $-1/2$ & $0$ & $1$ & $\alpha_{2}$ & $\alpha_{2}$ &
$\alpha_{2}-1$ & $+1/4$ & $-3/4$ & $+1/4$\\\hline
\end{tabular}
\end{equation}
or%
\begin{equation}%
\begin{tabular}
[c]{cccccccccc}\hline
$\bar{\nu}_{R}\mathbf{Y}_{\nu}\ell_{L}\Phi_{2}$ & $\phi$ & $\Phi_{1}$ &
$\Phi_{2}$ & $q_{L}$ & $u_{R}$ & $d_{R}$ & $\ell_{L}$ & $e_{R}$ & $\nu_{R}%
$\\\hline
$\ \ U(1)_{1}\ \ $ & $+1/2$ & $1$ & $0$ & $\alpha_{1}$ & $\alpha_{1}+1$ &
$\alpha_{1}$ & $-1/4$ & $-1/4$ & $-1/4$\\
$\ \ U(1)_{2}\ \ $ & $-1/2$ & $0$ & $1$ & $\alpha_{2}$ & $\alpha_{2}$ &
$\alpha_{2}-1$ & $-3/4$ & $-7/4$ & $+1/4$\\\hline
\end{tabular}
\end{equation}
To ensure that $U(1)_{Y}\subset U(1)_{1}\otimes U(1)_{2}$, we must set
$\alpha_{1}+\alpha_{2}=1/3$, and the remaining one-parameter freedom
originates in the $\mathcal{B}$ invariance of the Yukawa couplings. Yet,
clearly, no linear combination of the $U(1)_{1}$ and $U(1)_{2}$ charges can
make the Higgs doublets and the quarks neutral. So $U(1)_{\mathcal{L}}\subset U(1)_{1}\otimes U(1)_{2}$ does not correspond to the usual lepton number.

The symmetry breaking proceeds as in the DFSZ model since the scalar potential
stays the same. This fixes the PQ charge of the scalar fields to the same
values, Eq.~(\ref{PQdfsz}).\ The fermions then have the same charge as in
Eq.~(\ref{PQferm}), but with $\beta$ fixed so that $PQ(\nu_{R})=-PQ(\phi)/2$:%
\begin{subequations}
\begin{align}
\bar{\nu}_{R}\mathbf{Y}_{\nu}\ell_{L}\Phi_{1}  &  \Rightarrow\beta=-\frac
{1}{4}\left(  5x+\frac{1}{x}\right) \nonumber\\
&  \Rightarrow PQ(\ell_{L},e_{R},\nu_{R})=-\frac{1}{4}\left(  5x+\frac{1}%
{x},5x-\frac{3}{x},x+\frac{1}{x}\right)  \ \ ,\label{betanuDFSZ1}\\
\bar{\nu}_{R}\mathbf{Y}_{\nu}\ell_{L}\Phi_{2}  &  \Rightarrow\beta=-\frac
{1}{4}\left(  x-\frac{3}{x}\right) \nonumber\\
&  \Rightarrow PQ(\ell_{L},e_{R},\nu_{R})=-\frac{1}{4}\left(  x-\frac{3}%
{x},x-\frac{7}{x},x+\frac{1}{x}\right)  \ , \label{betanuDFSZ2}%
\end{align}
together with $PQ(q_{L},u_{R},d_{R})=\left(  \alpha,\alpha+x,\alpha
+1/x\right)  $, as before. In some sense, $U(1)_{\mathcal{L}}$ never occurs at
low energy. Instead, it is embedded into $U(1)_{PQ}$ via the specific value of
$\beta$ imposed by the $\bar{\nu}_{R}^{C}\mathbf{Y}_{R}\nu_{R}\phi$ coupling.
So, in this model, the axion and majoron are really one and the same particle.
Further, the \textquotedblleft axion~=~majoron\textquotedblright\ is
automatically coupled to quarks and to $G_{\mu\nu}^{a}\tilde{G}^{a,\mu\nu}$,
hence can solve the strong CP puzzle via the same mechanism as in the DFSZ model.

Once $\phi$ acquires its vacuum expectation value, $\nu_{R}$ has a Majorana
mass term, so it may seem this contradicts the fact that $PQ(\nu_{R})\neq0$.
But actually, plugging Eq.~(\ref{PolPhi}) in $\mathcal{L}_{\nu_{R}}$ of
Eq.~(\ref{SingletSeesaw}) and using Eq.~(\ref{RotDFSZ}), we find%
\end{subequations}
\begin{equation}
\bar{\nu}_{R}^{C}\mathbf{Y}_{R}\nu_{R}\phi\rightarrow\bar{\nu}_{R}%
^{C}\mathbf{M}_{R}\nu_{R}\exp(i\eta_{s}/v_{s})\approx\bar{\nu}_{R}%
^{C}\mathbf{M}_{R}\nu_{R}\exp(iPQ(\phi)a^{0}/v_{s})\ , \label{PhiMajo}%
\end{equation}
where $\mathbf{M}_{R}=v_{s}\mathbf{Y}_{R}/\sqrt{2}$. Thus, we now see why
$\nu_{R}$ must have a non-zero PQ charge. Because $\mathbf{M}_{R}$ orginates
from the $\phi$ field, it is always accompanied by the axion field. Then,
under a $U(1)_{PQ}$ transformation, $a\rightarrow a+v_{s}\theta$ must be
compensated by the phase shift $\nu_{R}\rightarrow\nu_{R}\exp(iPQ(\nu
_{R})\theta)$. Also, thanks to this, the fermion field reparametrization
$\psi\rightarrow\psi\exp(iPQ(\psi)a/v_{s})$ are still able to entirely move
the axion field out of the fermion mass terms.

One point must be stressed though. The axion couples to SM fermions via its
suppressed components $\eta_{1,2}$, but it couples directly to $\nu_{R}$ via
its dominant $\eta_{s}$ component. As a result, the couplings to SM fermions
are $\mathcal{O}(v/v_{s})$, but that to $\nu_{R}$ is $\mathcal{O}(1)$, as
evident in Eq.~(\ref{PhiMajo}). Yet, since $v_{s}$ is assumed to be well above
the electroweak scale, $\nu_{R}$ should be integrated out before performing
the fermion reparametrization.\ In that case, we find (assuming $\mathbf{Y}%
_{R}=\mathbf{Y}_{R}^{\dagger}$)
\begin{equation}
\mathcal{L}_{seesaw}^{eff}=-\frac{1}{2}(\bar{\ell}_{L}^{C}\Phi_{i}%
^{T})\mathbf{Y}_{\nu}^{T}\frac{1}{\mathbf{M}_{R}}\mathbf{Y}_{\nu}(\ell_{L}%
\Phi_{i})\exp(i(2PQ(\Phi_{i})-2PQ(\nu_{R}))a^{0}/v_{s})+h.c.
\end{equation}
Then, performing $\ell_{L}\rightarrow\ell_{L}\exp(iPQ(\ell_{L})a/v_{s})$ moves
the axion field entirely into the same effective derivative and anomalous
interactions as in Eq.~(\ref{dJPQTHDM}), but with $\beta$ now fixed as in
Eq.~(\ref{betanuDFSZ1}) or~(\ref{betanuDFSZ2}). Again, the phenomenology is
unaffected since $\beta$ cancels out of physical observables. Thus, the
$\mathcal{O}(1)$ axion coupling to $\nu_{R}$ has no consequences at low
energy.

\subsection{PQ and DFSZ with a type II seesaw mechanism}

In the previous sections, we have seen two ways to incorporate neutrino masses
in the DFSZ model.\ For the first, one simply adds right handed neutrinos
$\nu_{R}$ with a Majorana mass term. The PQ symmetry stays the same, though a
specific value of $\beta$ is required, Eq.(\ref{betaPQ}), to ensure
$PQ(\nu_{R})=0$. Also, this makes sure the explicit breaking of the lepton
number does not spill over to the PQ symmetry. A second way to proceed, in the
$\nu$DFSZ model, is again to add right-handed neutrinos but ask to the heavy
singlet field to induce their Majorana mass term.\ In that case, $PQ(\nu
_{R})\neq0$, but the lepton number symmetry ceased to exist. Actually, it is
replaced by the PQ symmetry.

A third realization is provided by the type II seesaw mechanism~\cite{TypeII}.
Instead of right-handed neutrinos, let us add to the THDM model three complex
Higgs fields $\mathbf{\Delta}$ transforming as a $SU(2)_{L}$ triplet with
hypercharge 2. For the couplings to fermions, in addition to the THDM Yukawas,
we add to $\mathcal{L}_{\text{Yukawa}}$ of Eq.~(\ref{YukQuark}) the term%
\begin{equation}
\mathcal{L}_{\nu II}=\bar{\ell}_{L}^{C}Y_{\Delta}\mathbf{\Delta}\ell
_{L}\ ,\ \ \ell_{L}^{C}=i\sigma^{2}\left(
\begin{array}
[c]{c}%
\nu_{L}^{C}\\
\ell_{L}^{C}%
\end{array}
\right)  \ ,\ \ \mathbf{\Delta}=\frac{1}{\sqrt{2}}\left(
\begin{array}
[c]{cc}%
\Delta^{+} & \sqrt{2}\Delta^{++}\\
\Delta^{0} & -\Delta^{+}%
\end{array}
\right)  \ ,
\end{equation}
where, as indicated, $C$ acts in both Lorentz and $SU(2)_{L}$ spaces. For the
scalar potential, we introduce one new coupling to entangle the $U(1)_{1}%
\otimes U(1)_{2}$ charges of $\mathbf{\Delta}$ with those of the doublets, so
that $V_{\nu2\text{THDM}}=V_{\text{THDM}}+V_{\Delta}+V_{\Delta\text{THDM}%
}+V_{\Delta\text{PQ}}$ with%
\begin{subequations}
\begin{align}
V_{\Delta}  &  =\mu_{\Delta}^{2}\langle\mathbf{\Delta}^{\dagger}%
\mathbf{\Delta}\rangle+\lambda_{\Delta1}\langle\mathbf{\Delta}^{\dagger
}\mathbf{\Delta}\rangle^{2}+\lambda_{\Delta2}\langle(\mathbf{\Delta}^{\dagger
}\mathbf{\Delta})^{2}\rangle\ ,\frac{{}}{{}}\\
V_{\Delta\text{THDM}}  &  =a_{\Delta1}\langle\mathbf{\Delta}^{\dagger
}\mathbf{\Delta}\rangle\Phi_{1}^{\dagger}\Phi_{1}+a_{\Delta2}\langle
\mathbf{\Delta}^{\dagger}\mathbf{\Delta}\rangle\Phi_{2}^{\dagger}\Phi
_{2}+a_{\Delta3}\Phi_{1}^{\dagger}\mathbf{\Delta\Delta}^{\dagger}\Phi
_{1}+a_{\Delta4}\Phi_{2}^{\dagger}\mathbf{\Delta\Delta}^{\dagger}\Phi
_{2}\ ,\frac{{}}{{}}\\
V_{\Delta\text{PQ}}  &  =-\mu_{\Delta}\lambda_{\Delta12}\tilde{\Phi}_{1}%
^{T}\Delta^{\dagger}\Phi_{2}+h.c.\;,\frac{{}}{{}}%
\end{align}
with $\tilde{\Phi}_{i}=i\sigma^{2}\Phi_{i}$. A factor $\mu_{\Delta}$ is
introduced to make $\lambda_{\Delta12}$ dimensionless. Even if $\mu_{\Delta
}^{2}$ is large and positive, the $\lambda_{\Delta12}$ coupling generates a
tadpole for $\operatorname{Re}\Delta^{0}$ and this field has to be shifted. In
effect, this induces a VEV for the $\Delta$ field, $v_{\Delta}\sim
\lambda_{\Delta12}v_{1}v_{2}/\mu_{\Delta}$. To preserve the electroweak
custodial symmetry, $\mu_{\Delta}\gg v_{1,2}$ so that $v_{\Delta}\ll v_{1,2}$.
Yet, this shift generates a small Majorana mass term for the neutrinos,
$m_{\nu}=v_{\Delta}Y_{\Delta}$. This is the characteristic linear suppression
of the neutrino masses of the type II seesaw mechanism.

We have not identified the axion field yet. To that end, we adopt again the
polar parametrization, Eq.~(\ref{PolPara}) together with%
\end{subequations}
\begin{equation}
\mathbf{\Delta}=\frac{1}{\sqrt{2}}\exp(i\eta_{\Delta}/v_{\Delta})\left(
\begin{array}
[c]{cc}%
\Delta^{+} & \sqrt{2}\Delta^{++}\\
v_{\Delta}+\operatorname{Re}\Delta^{0} & -\Delta^{+}%
\end{array}
\right)  \ .
\end{equation}
Restricted to the pseudoscalar states, only the $\lambda_{\Delta12}$ coupling
survives and%
\begin{equation}
V_{\nu2\text{THDM}}(\eta_{1,2,\Delta})=-\frac{1}{\sqrt{2}}\mu_{\Delta}%
\lambda_{\Delta12}v_{1}v_{2}v_{\Delta}\cos\left(  \frac{\eta_{1}}{v_{1}}%
+\frac{\eta_{2}}{v_{2}}-\frac{\eta_{\Delta}}{v_{\Delta}}\right)  \ .
\label{CosTypII}%
\end{equation}
The mass eigenstates are easily found. First, the $G^{0}$ state has to be
aligned with%
\begin{equation}
G^{0}\sim v_{1}\eta_{1}+v_{2}\eta_{2}+2v_{\Delta}\eta_{\Delta}\ ,
\label{GOsII}%
\end{equation}
since this ensures it can be removed by a $U(1)_{Y}$ transformation and
$Y(\Delta)=2Y(\Phi_{1,2})=2$. Second, the single massive state, denoted
$\pi^{0}$, is aligned with the combination of fields occurring in the argument
of the cosine in Eq.~(\ref{CosTypII}). The axion is then the only state
orthogonal to both $G^{0}$ and $\pi^{0}$, and a simple cross product permits
to construct the mixing matrix:%
\begin{equation}
\left(
\begin{array}
[c]{c}%
G^{0}\\
a^{0}\\
\pi^{0}%
\end{array}
\right)  =\left(
\begin{array}
[c]{ccc}%
s_{\beta}\omega & c_{\beta}\omega & 2\delta_{\Delta}\omega\\
c_{\beta}\left(  1+2\delta_{\Delta}^{2}/c_{\beta}^{2}\right)  \omega
\omega^{\prime} & -s_{\beta}\left(  1+2\delta_{\Delta}^{2}/s_{\beta}%
^{2}\right)  \omega\omega^{\prime} & 2\delta_{\Delta}\omega\omega^{\prime
}/t_{2\beta}\\
\delta_{\Delta}\omega^{\prime}/s_{\beta} & \delta_{\Delta}\omega^{\prime
}/c_{\beta} & -\omega^{\prime}%
\end{array}
\right)  \left(
\begin{array}
[c]{c}%
\eta_{1}\\
\eta_{2}\\
\eta_{\Delta}%
\end{array}
\right)  \ ,
\end{equation}
with $\delta_{\Delta}=v_{\Delta}/v$, $\omega^{-2}=1+4\delta_{\Delta}^{2}$ and
$\omega^{\prime-2}=1+4\delta_{\Delta}^{2}/s_{2\beta}^{2}$. The PQ charges of
the three fields can be read off the second line of this matrix, and upon
adopting a convenient normalization:%
\begin{equation}
PQ(\Phi_{1},\Phi_{2},\Delta)=\left(  x+2x_{\Delta},\ -\frac{1}{x}-2x_{\Delta
}\ ,\ x-\frac{1}{x}\right)  \ ,\ \ x_{\Delta}\equiv\delta_{\Delta}^{2}\left(
x+\frac{1}{x}\right)  \ . \label{PQSIIscalar}%
\end{equation}
Note that $PQ(\Phi_{1})+PQ(\Phi_{2})-PQ(\Delta)=0$, as it should, but those
can be expressed as simple function of $x=1/\tan\beta$ only to leading order
in $\delta_{\Delta}$.

Once the PQ charge of the scalars is set, that of the fermions can be derived
and we find%
\begin{equation}
PQ(q_{L},u_{R},d_{R},\ell_{L},e_{R})=\left(  \alpha,\alpha+x+2x_{\Delta
},\alpha+\frac{1}{x}+2x_{\Delta},-\frac{x}{2}+\frac{1}{2x},-\frac{x}{2}%
+\frac{3}{2x}+2x_{\Delta}\right)  \ \ . \label{PQSIIferm}%
\end{equation}
Apart from the small shifts induced by $x_{\Delta}$, this corresponds to the
PQ current of the THDM with $\beta=-PQ(\Delta)/2$.

Since the $\eta_{1,2}$ components of $a^{0}$ are of $\mathcal{O}(1)$, the PQ
scale stays at $v$, and the axion ends up too strongly coupled to SM fermions.
To cure for this, the same strategy as in the DFSZ model can be used, that is,
an additional complex singlet field is
introduced~\cite{Bertolini:1990vz,Bertolini:2014aia}.\ To study this
situation, let us take the scalar potential%
\begin{align}
V_{\nu2\text{DFSZ}}  &  =V_{\text{THDM}}+V_{\Delta}+V_{\Delta\text{THDM}%
}+V_{\phi}+V_{\phi\text{THDM}}+b_{\phi\Delta}\phi^{\dagger}\phi\langle
\mathbf{\Delta}^{\dagger}\mathbf{\Delta}\rangle\nonumber\\
&  +\left[  -\lambda_{\nu1}\phi^{2}\Phi_{1}^{\dagger}\Phi_{2}-\lambda_{\nu
2}\phi\Phi_{1}^{T}\Delta^{\dagger}\Phi_{2}-\mu_{\Delta}\lambda_{\nu3}\Phi
_{1}^{T}\Delta^{\dagger}\Phi_{2}+h.c.\right]  \ .
\end{align}
The coupling $b_{\phi\Delta}$ gives a large $\mathcal{O}(v_{s})$ mass to the
triplet states, while those in $V_{\Delta\text{THDM}}$ generate small
$\mathcal{O}(v)$ splittings among the three $\Delta$ states. Of particular
interests are the $\lambda_{\nu i}$ couplings since they entangle the scalar
states. First, the $\lambda_{\nu2}$ and $\lambda_{\nu3}$ couplings creates
$\Delta$ tadpoles that need to be removed by shifting the $\Delta$ field%
\begin{equation}
v_{\Delta}=\frac{1}{2}\lambda_{\nu2}\frac{v_{1}v_{2}v_{s}}{\mu_{\Delta}%
^{2}+b_{\phi\Delta}v_{s}^{2}}+\frac{1}{2}\lambda_{\nu3}\frac{\mu_{\Delta}%
v_{1}v_{2}}{\mu_{\Delta}^{2}+b_{\phi\Delta}v_{s}^{2}}\ .
\end{equation}
Note that if $\mu_{\Delta}\ll v_{s}$, the bulk of the $\Delta$ mass comes from
the singlet, and the $\mu_{\Delta}^{2}$ can be neglected in these expressions.

Plugging in the polar representations of the scalar fields, the scalar
potential for the pseudoscalar states is:%
\begin{align}
V_{\nu2\text{DFSZ}}(\eta_{1,2,\Delta,s})  &  =-\frac{1}{2}\lambda_{\nu1}%
v_{1}v_{2}v_{s}^{2}\cos\left(  \frac{\eta_{1}}{v_{1}}-\frac{\eta_{2}}{v_{2}%
}-\frac{2\eta_{s}}{v_{s}}\right) \nonumber\\
&  \ \ \ \ -\frac{1}{2}\lambda_{\nu2}v_{1}v_{2}v_{s}v_{\Delta}\cos\left(
\frac{\eta_{1}}{v_{1}}+\frac{\eta_{2}}{v_{2}}+\frac{\eta_{s}}{v_{s}}%
-\frac{\eta_{\Delta}}{v_{\Delta}}\right) \nonumber\\
&  \ \ \ \ -\frac{1}{\sqrt{2}}\mu_{\Delta}\lambda_{\nu3}v_{1}v_{2}v_{\Delta
}\cos\left(  \frac{\eta_{1}}{v_{1}}+\frac{\eta_{2}}{v_{2}}-\frac{\eta_{\Delta
}}{v_{\Delta}}\right)  \ .
\end{align}
If the three $\lambda_{\nu i}$ couplings are present, there are three massive
pseudoscalar states corresponding to the linear combinations appearing in the
cosine functions. Those are linearly independent. Together with $G^{0}$ which
stays of course massless, there is no room for the axion. This was evident
from the start, since with all three $\lambda_{\nu i}$ couplings, no
$U(1)_{1}\otimes U(1)_{2}$ symmetry can be defined.

Removing any one of these couplings, a second Goldstone boson appears and can
be identified with the axion. Given that the $G^{0}$ state stays the same as
without the $\phi$, see Eq.~(\ref{GOsII}), we directly find the $a^{0}$ state
by its orthogonality with $G^{0}$ and the massive states in the cosine
functions,%
\begin{align}
\lambda_{\nu1}  &  =0:a^{0}\sim(v_{2}^{2}+2v_{\Delta}^{2})v_{1}\eta_{1}%
-(v_{1}^{2}+2v_{\Delta}^{2})v_{2}\eta_{2}+(v_{2}^{2}-v_{1}^{2})v_{\Delta}%
\eta_{\Delta}\ ,\\
\lambda_{\nu2}  &  =0:a^{0}\sim2(v_{2}^{2}+2v_{\Delta}^{2})v_{1}\eta
_{1}-2(v_{1}^{2}+2v_{\Delta}^{2})v_{2}\eta_{2}+2(v_{2}^{2}-v_{1}^{2}%
)v_{\Delta}\eta_{\Delta}+(v_{1}^{2}+v_{2}^{2}+4v_{\Delta}^{2})v_{s}\eta
_{s}\ ,\\
\lambda_{\nu3}  &  =0:a^{0}\sim2(v_{2}^{2}+v_{\Delta}^{2})v_{1}\eta
_{1}-2(v_{1}^{2}+3v_{\Delta}^{2})v_{2}\eta_{2}+(3v_{2}^{2}-v_{1}^{2}%
)v_{\Delta}\eta_{\Delta}+(v_{1}^{2}+v_{2}^{2}+4v_{\Delta}^{2})v_{s}\eta
_{s}\ \ .
\end{align}

The first scenario collaspes to that without $\phi$, and is ruled out since
the axion scale remains at $v$.\ The other two are viable, with the axion
scale set by $v_{s}$.\ The PQ assignments can be read off the coefficients of
the $v_{i}\eta_{i}$ terms above, and upon adopting a convenient
normalization,
\begin{align}
\lambda_{\nu2} &  =0:PQ(\Phi_{1},\Phi_{2},\Delta,\phi)=\left(  x+2x_{\Delta
},\ -\frac{1}{x}-2x_{\Delta}\ ,\ x-\frac{1}{x}\ ,\ \frac{x}{2}+\frac{1}%
{2x}+2x_{\Delta}\right)  \ ,\\
\lambda_{\nu3} &  =0:PQ(\Phi_{1},\Phi_{2},\Delta,\phi)=\left(  x+x_{\Delta
},\ -\frac{1}{x}-3x_{\Delta}\ ,\ \frac{3}{2}x-\frac{1}{2x}\ ,\ \frac{x}%
{2}+\frac{1}{2x}+2x_{\Delta}\right)  \ ,
\end{align}
with $x_{\Delta}$ given in Eq.~(\ref{PQSIIscalar}). The $\lambda_{\nu2}=0$
scenario is the simple DFSZ generalization of the THDM with a type II seesaw,
and the PQ charges stay the same, see Eq.~(\ref{PQSIIscalar}). Consequently,
the fermions have the charges in Eq.~(\ref{PQSIIferm}).

For the $\lambda_{\nu3}=0$ scenario, corresponding to that discussed in
Ref.~\cite{Bertolini:1990vz,Bertolini:2014aia}, the PQ charges of the leptons
are shifted since that of $\Delta$ is different:%
\begin{equation}
PQ(q_{L},u_{R},d_{R},\ell_{L},e_{R})=\left(  \alpha,\alpha+x+x_{\Delta}%
,\alpha+\frac{1}{x}+3x_{\Delta},-\frac{3}{4}x+\frac{1}{4x},-\frac{3}{4}%
x+\frac{5}{4x}+3x_{\Delta}\right)  \ \ .
\end{equation}
Again, apart from the small shifts induced by $v_{\Delta}$, the PQ charges in
these two scenarios correspond to that of the THDM in Eq.~(\ref{PQferm}) with
specific values of $\beta$:%
\begin{align}
\lambda_{\nu2}  &  =0\Rightarrow\beta=-\frac{3}{4}x+\frac{1}{4x}\ ,\\
\lambda_{\nu3}  &  =0\Rightarrow\beta=-\frac{x}{2}+\frac{1}{2x}\ .
\end{align}
The electroweak terms in the divergence of the PQ current are thus different
in both scenarios.\ Yet, phenomenologically, the axion couplings are
independent of $\beta$, and apart from negligible corrections brought in by
$v_{\Delta}$, these scenarios cannot be distinguished at low energy.

\section{Axions and baryon number violation\label{sec4}}

Up to now, we have seen that the violation of the lepton number, through
insertion of Majorana neutrino masses, fixes one of the two ambiguities in the
PQ charges of the SM fermions, that parametrized by $\beta$ in
Eq.~(\ref{PQferm}). We will now concentrate on the remaining ambiguity,
$\alpha$, which originates in the conserved baryon number current. In the
first subsection, we will discuss two frameworks in which $\alpha$ is
automatically fixed, for dynamical reasons. Then, in the second subsection,
the impact of explicit B-violating operators will be discussed. Finally, in
the last subsection, the situation in which too much $\mathcal{B}$ and/or
$\mathcal{L}$ violating effects are introduced, preventing the PQ symmetry
from arising, will be described.

\subsection{Dynamical $\mathcal{B}$ violation}

Even without explicit $\mathcal{B}$ violation, $U(1)_{\mathcal{B}}$ is not a
true symmetry at the quantum level because electroweak instantons are known to
induce $\mathcal{B}+\mathcal{L}$
transitions~\cite{tHooft:1976snw,tHooft:1976rip}. This takes the form of an
effective interaction involving antisymmetric flavor contractions of three
lepton weak doublets and nine quark weak doublets:
\begin{equation}
\mathcal{L}_{inst}^{eff}=c_{inst}\ell_{L}^{3}q_{L}^{9}\ .
\end{equation}
At zero temperature, $c_{inst}$ is tuned by $\exp(-4\pi/g^{2})$ and these
effects are totally negligible. Yet, even so, these interactions are present,
and following the same philosophy as for $\beta$, they prevent the emergence
of the parametric freedom to choose $\alpha$ and $\beta$ separately.
Specifically, the PQ symmetry necessarily settles with%
\begin{equation}
\Delta\mathcal{B}=\Delta\mathcal{L}=3\quad\Rightarrow\quad3\alpha+\beta=0\ .
\label{abInst}%
\end{equation}
Setting this combination to zero also kills off the $W_{\mu\nu}^{i}\tilde
{W}^{i,\mu\nu}$ term in the PQ current (see Eq.~(\ref{dJPQTHDM})). In some
sense, this requirement removes a $\mathcal{B}+\mathcal{L}$ component in
$U(1)_{PQ}$. Since $\mathcal{B}-\mathcal{L}$ is anomaly-free, there is then
nothing remaining to generate the $W_{\mu\nu}^{i}\tilde{W}^{i,\mu\nu}$ term.
Yet, there remain couplings of the axion to left-handed fields in the
effective non-linear Lagrangian since neither $\alpha$ nor $\beta$ are
vanishing when Majorana neutrino masses are present.

It should be mentionned also that once the electroweak instanton interaction
fixes $3\alpha+\beta=0$, the axion decouples from electroweak anomalous
effects, including also the sphaleron interactions. Mechanisms to generate the
baryon asymmetry from the rotation of an axion field (see e.g.
Ref.~\cite{Co:2019wyp}) which relies on those interactions cannot be active in
the present simple axion models. Some additional constraints must force the PQ
symmetry to be realized differently. At the very least, it must fix $\alpha$
to some value not compatible with the $\beta$ value imposed by the neutrino
sector, in order to induce $3\alpha+\beta\neq0$.

So, generically, electroweak instantons prevent the emergence of one of the
ambiguities in the fermionic PQ charges. However, this supposes the ambiguity
is not removed first at a yet higher scale. A generic class of models where
this occurs are the GUT scenarios. Indeed, in that case, gauge interactions
can break $\mathcal{B}$ and $\mathcal{L}$. For example, in $SU(5)$,
$\mathcal{B}-\mathcal{L}$ is conserved but not $\mathcal{B}+\mathcal{L}$, and
one automatically has%
\begin{equation}
\text{GUT}:\quad3\alpha+\beta=-x-\frac{1}{x}\ . \label{abGUT}%
\end{equation}
Indeed, this is the unique value for which all three anomalous terms in
Eq.~(\ref{JPQTHDM}) coincide when taking into account the $SU(5)$
normalization of the hypercharge, $\mathcal{N}_{C}=\mathcal{N}_{L}%
=5/3\mathcal{N}_{Y}$. It is quite remarkable that this value is not compatible
with the instanton value in Eq.~(\ref{abInst}). Further investigation of the
fermion charge ambiguities in a GUT context are defered to future work~\cite{toappear}.

\subsection{Effective $\mathcal{B}$ violation}

The basis of effective, gauge invariant operators violating $\mathcal{B}$
and/or $\mathcal{L}$ is well-known. It starts at the dimension-five level with
the $\Delta\mathcal{L}=2$ Majorana mass operator of Eq.~(\ref{leffseesaw}).
Then, at the dimension-six level, all the operators are $\Delta\mathcal{B}%
=\Delta\mathcal{L}=1$ (see Ref.~\cite{Weinberg:1980bf}):%
\begin{equation}
\mathcal{L}_{eff}^{\dim6}=\frac{1}{\Lambda^{2}}(\ell_{L}q_{L}^{3}+e_{R}%
u_{R}^{2}d_{R}+e_{R}u_{R}q_{L}^{2}+\ell_{L}q_{L}d_{R}u_{R})+h.c.\ .
\end{equation}
Adequate contractions of the Lorentz and $SU(2)_{L}$ spinors are understood, as
well as Wilson coefficients and flavor indices. Beyond that level, other
patterns of $\Delta\mathcal{B}$ and $\Delta\mathcal{L}$ can occur at the
dimension-seven level, thanks to additional Higgs insertions. With only SM fermions, the next series of operators arise at the dimension-nine level:%
\begin{equation}
\mathcal{L}_{eff}^{\dim9}=\frac{1}{\Lambda^{5}}(e_{R}\ell_{L}^{2}u_{R}%
^{3}+\ell_{L}^{3}q_{L}u_{R}^{2}+d_{R}^{4}u_{R}^{2}+d_{R}^{3}u_{R}q_{L}%
^{2}+d_{R}^{2}q_{L}^{4})+h.c.\ .
\end{equation}
The first two induce $\Delta\mathcal{B}=1,\Delta\mathcal{L}=3$ transitions,
and the last three $\Delta\mathcal{B}=2,\Delta\mathcal{L}=0$ ones. These
operators are peculiar because, provided the flavor indices are
antisymmetrically contracted, they break only $U(1)_{\mathcal{L}}$ and
$U(1)_{\mathcal{B}}$ and not the flavor $SU(3)$s~\cite{Smith:2011rp}.

Given the charges in Eq.~(\ref{PQferm}), none of these operators is invariant
under $U(1)_{PQ}$, but carry instead
\begin{subequations}
\label{PQdim6}%
\begin{align}
PQ(\ell_{L}q_{L}^{3})  &  =3\alpha+\beta\ ,\\
PQ(e_{R}u_{R}^{2}d_{R})  &  =3\alpha+\beta+2x+\frac{2}{x}\ ,\\
PQ(e_{R}u_{R}q_{L}^{2})  &  =3\alpha+\beta+x+\frac{1}{x}\ ,\\
PQ(\ell_{L}q_{L}d_{R}u_{R})  &  =3\alpha+\beta+x+\frac{1}{x}\ .
\end{align}
and
\end{subequations}
\begin{subequations}
\begin{align}
PQ(e_{R}\ell_{L}^{2}u_{R}^{3})  &  =3\alpha+3\beta+3x+\frac{1}{x}\ ,\\
PQ(\ell_{L}^{3}q_{L}u_{R}^{2})  &  =3\alpha+3\beta+2x\ ,\\
PQ(d_{R}^{4}u_{R}^{2})  &  =6\alpha+2x+\frac{4}{x}\ ,\\
PQ(d_{R}^{3}u_{R}q_{L}^{2})  &  =6\alpha+x+\frac{3}{x}\ ,\\
PQ(d_{R}^{2}q_{L}^{4})  &  =6\alpha+\frac{2}{x}\ .
\end{align}
The way in which $\alpha$ and $\beta$ enters reflects the $\Delta\mathcal{B}$
and $\Delta\mathcal{L}$ properties of the corresponding operators, with
$n\times\alpha\Leftrightarrow\Delta\mathcal{B}=n/3$ and $m\times
\beta\Leftrightarrow\Delta\mathcal{L}=m$. Yet, remarkably, the PQ charge of
the operators are not aligned with their $\Delta\mathcal{B}$ and
$\Delta\mathcal{L}$ contents. For example, all the dimension-six operators are
$\Delta\mathcal{B}=\Delta\mathcal{L}=1$, but they do have different PQ charge.
Among the dimension-six operators, it is also interesting to remark that only
the first is compatible with electroweak instantons, Eq.~(\ref{abInst}), while
only the last two are compatible with GUTs, Eq.~(\ref{abGUT}). This could have
been expected since those are the operators arising from $SU(5)$ gauge boson exchanges.

Another noticeable feature is that the misalignment between two operators
carrying the same $\Delta\mathcal{B}$ and $\Delta\mathcal{L}$ always appears
as a mutliple of $x+1/x$. This means that even if the PQ symmetry does not
exist when all types of operators are simultaneously present, large classes of
operators can nevertheless be allowed, but at the cost of a further scaling in
their dimensions. Consider for instance the DFSZ model where $PQ(\Phi
_{2}^{\dagger}\Phi_{1})=PQ(\phi^{\dagger})=x+1/x$. Misalignments can always be
compensated by scalar singlet insertions. For instance, if one assumes
Eq.~(\ref{abInst}) holds, then the $\Delta\mathcal{B}=\Delta\mathcal{L}=1$
operators must be%
\end{subequations}
\begin{equation}
\mathcal{L}_{eff}^{\Delta(\mathcal{B}-\mathcal{L})=0}=\frac{\ell_{L}q_{L}^{3}%
}{\Lambda^{2}}+\frac{e_{R}u_{R}q_{L}^{2}}{\Lambda^{3}}\phi+\frac{\ell_{L}%
q_{L}d_{R}u_{R}}{\Lambda^{3}}\phi+\frac{e_{R}u_{R}^{2}d_{R}}{\Lambda^{4}}%
\phi^{2}+h.c.\ .
\end{equation}
The PQ charge of all the operators is now aligned in the direction of
$\ell_{L}q_{L}^{3}$, with $PQ(\ell_{L}q_{L}^{3})=0$ when $3\alpha+\beta=0$.
Operators involving insertions of the Higgs doublet combination $\Phi
_{1}^{\dagger}\Phi_{2}$ need not be included since\ they are comparatively
very suppressed, both dimensionally and because $v_{1,2}\ll v_{s}$.
Phenomenologically, given the bounds on $\Lambda$ from proton decay and
provided $v_{s}<\Lambda$, only the leading operator is expected to play any
role. The same holds for other series of operators, though there is then no
clear reason to select one operator against another as leading.\ For example,
in the $\Delta\mathcal{B}=2$ class, assuming $d_{R}^{4}u_{R}^{2}$ is leading,
the effective operators must be
\begin{equation}
\mathcal{L}_{eff}^{\Delta\mathcal{B}=2}=\frac{d_{R}^{4}u_{R}^{2}}{\Lambda^{5}%
}+\frac{d_{R}^{3}u_{R}q_{L}^{2}}{\Lambda^{6}}\phi+\frac{d_{R}^{2}q_{L}^{4}%
}{\Lambda^{7}}\phi^{2}+h.c.\ .
\end{equation}
They are all neutral provided $3\alpha=-x-2/x$, and thus there remains enough
room for the PQ symmetry to exist.

\subsection{Vacuum realignments}

In the previous section, we have seen that the PQ symmetry can accommodate for
limited $\mathcal{B}$ and/or $\mathcal{L}$ breaking. Our goal here is to work
out the consequences when too much $\mathcal{B}$ and/or $\mathcal{L}$
violation is introduced. Indeed, if there are too many misaligned
$\Delta\mathcal{B}$ and $\Delta\mathcal{L}$ operators, the $U(1)_{1}\otimes
U(1)_{2}$ symmetry cannot be exact and the axion cannot be massless.

To analyze this, we first remark that these breaking effects have to be tiny
given the experimental constraints on $\Delta\mathcal{B}$ and $\Delta
\mathcal{L}$ transitions. Thus, the $U(1)_{1}\otimes U(1)_{2}$ symmetry is at
most only very slightly broken, and the leading dynamics remain that of
Goldstone bosons. The pseudoscalar degrees of freedom can still be
parametrized using the polar representation. Of course, the axion will no
longer be truly massless, it becomes a pseudo-Goldstone boson. Naively, if
this mass is too large compared to the QCD-induced mass, then the axion fails
to solve the strong CP puzzle. This failure can also be viewed in terms of the
vacuum of the theory. In the presence of the $\Delta\mathcal{B}$ and
$\Delta\mathcal{L}$ breaking terms, the shift symmetry is no longer active.
All the vacua are no longer equivalent, and one direction is prefered. At the
low-scale, QCD effects also require a realignment of the vacuum, and the
CP-puzzle can be solved only when the QCD requirement is stronger than that
coming from the $\Delta\mathcal{B}$ and $\Delta\mathcal{L}$ effects.

In the next section, the axion mass arising from various combinations of
$\Delta\mathcal{B}$ and $\Delta\mathcal{L}$ operators are analyzed
semi-quantitatively, from the point of view of the effective scalar potential.
Then, in the next section, we perform a more detailed analysis of the vacuum
realignment mechanism induced by the $\Delta\mathcal{B}$ and $\Delta
\mathcal{L}$ operators, in the spirit of Dashen theorem~\cite{Dashen:1970et}.

\subsubsection{Effective potential approach}

To estimate the mass of the axion, the simplest strategy is to start at the
level of the scalar potential before the electroweak SSB. Indeed, at tree
level, the $U(1)_{1}\otimes U(1)_{2}$ symmetry is still active there since it
is broken explicitly in the fermion sector only. Thus, at tree-level, the
axion remains as a massless Goldstone boson. To go beyond that, we must
consider the effective scalar potential, and in particular look for the
leading symmetry breaking terms induced by fermion loops. Clearly, such loops
must include all the misaligned $\Delta\mathcal{B}$ and/or $\Delta\mathcal{L}$
interactions simultaneously, in such a way that the process is $\Delta
\mathcal{B}=\Delta\mathcal{L}=0$ overall since scalar fields have
$\mathcal{B}=\mathcal{L}=0$.

\paragraph{Scenario I: Weinberg dimension-six operators and the axion mass.}

As a first situation, we consider the case where several operators inducing
the same $\Delta\mathcal{B}$ and $\Delta\mathcal{L}$ transitions are
introduced simultaneously. As discussed before, such a set of operators can be
organized into classes according to their PQ charges, with the PQ charges of
two classes differing by some multiple of $x+1/x$. Because this is precisely
the charge of the $\Phi_{2}^{\dagger}\Phi_{1}$ combination, the combined
presence of two operators whose PQ charge differ by $n\times(x+1/x)$ generates
the correction%
\begin{equation}
V_{\mathcal{B},\mathcal{L}}^{eff}=-\frac{2^{2n-3}\lambda_{n}}{\Lambda
_{\mathcal{B},\mathcal{L}}^{2n-4}}(\Phi_{2}^{\dagger}\Phi_{1})^{n}+h.c.\ ,
\label{VBLeff}%
\end{equation}
in the effective scalar potential, where $\Lambda_{\mathcal{B},\mathcal{L}}$
is the scale of the $\Delta\mathcal{B}$ and $\Delta\mathcal{L}$ physics,
$\lambda_{n}$ a complicated combination of the Wilson coefficients, Yukawa
couplings, and loop factors, and the factor $2^{2n-3}$ is introduced for
convenience. From there, the mass of the axion can be estimated as $m_{a^{0}%
}^{2}\sim\mathcal{O}(v^{2n-2}/\Lambda_{\mathcal{B},\mathcal{L}}^{2n-4})$ in
the PQ model.

This correction to the scalar potential is also valid for the DFSZ model,
since the singlet does not couple directly to fermions. The only way in which
the breaking of $U(1)_{1}\otimes U(1)_{2}$ can be communicated to $\phi$ is
via the mixing term $\phi^{2}\Phi_{1}^{\dagger}\Phi_{2}$. To incorporate this
effect, the full mass matrix for the pseudoscalar states has to be
diagonalized. To that end, consider the effective potential restricted to
pseudoscalar states. It now contains a second cosine function:%
\begin{equation}
V_{\text{DFSZ}}(\eta_{1,2,s})=-\frac{1}{2}\lambda_{12}v_{1}v_{2}v_{s}^{2}%
\cos\left(  \frac{\eta_{1}}{v_{1}}-\frac{\eta_{2}}{v_{2}}-\frac{2\eta_{s}%
}{v_{s}}\right)  -2^{n-1}\frac{\lambda_{n}(v_{1}v_{2})^{n}}{\Lambda
_{\mathcal{B},\mathcal{L}}^{2n-4}}\cos\left(  n\left(  \frac{\eta_{1}}{v_{1}%
}-\frac{\eta_{2}}{v_{2}}\right)  \right)  \ .
\end{equation}
Diagonalizing the mass matrix, one pseudoscalar state remains at the $v_{s}$
scale while the other has a mass%
\begin{equation}
m_{a^{0}}^{2}=\lambda_{n}n^{2}\frac{v^{2}}{v_{s}^{2}}\frac{v^{2n-2}}%
{\Lambda_{\mathcal{B},\mathcal{L}}^{2n-4}}\sin^{n}(2\beta)\ .
\label{GenericmA}%
\end{equation}
Thus, in the DFSZ model, the lightest pseudoscalar mass is suppressed by a
$v/v_{s}$ factor compared to the PQ model. Still, this factor does not really
help to make a scenario viable because the QCD contribution to the axion mass
also scales as $1/v_{s}$. For instance, with $m_{a^{0}}|_{QCD}\sim m_{\pi}%
^{2}/v_{s}$, $n$ should be strictly greater than two if $\lambda_{n}$ is
$\mathcal{O}(1)$ and $\Lambda_{\mathcal{B},\mathcal{L}}\approx10^{16}$~GeV.

\begin{figure}[t]
\centering\includegraphics[width=0.90\textwidth]{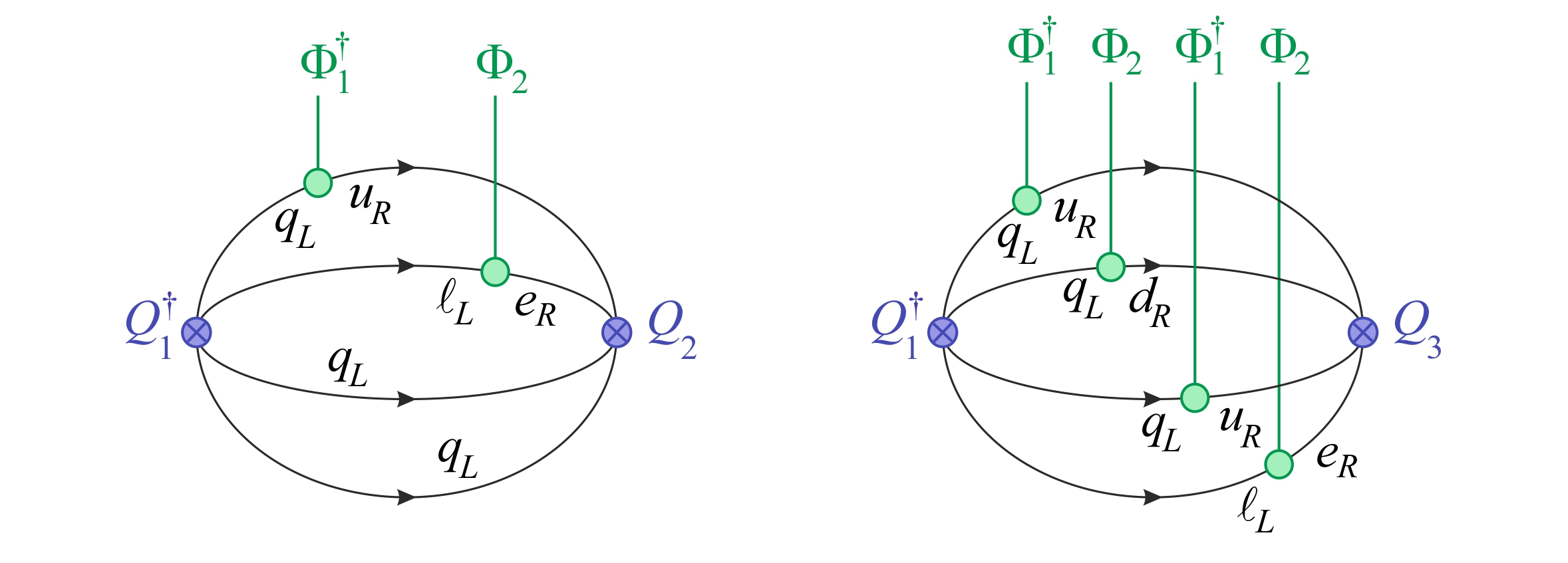}\caption{Fermion
loops involving the Weinberg operators $Q_{1}\equiv\ell_{L}q_{L}^{3}$,
$Q_{2}\equiv e_{R}u_{R}q_{L}^{2}$, and $Q_{3}\equiv e_{R}u_{R}^{2}d_{R}$, and
inducing symmetry-breaking effective potential terms.}%
\label{Fig1}%
\end{figure}

To illustrate this discussion, let us take the Weinberg operators of
Eq.~(\ref{PQdim6}). If both the $Q_{1}\equiv\ell_{L}q_{L}^{3}$ and
$Q_{2}\equiv e_{R}u_{R}q_{L}^{2}$ operators are simultaneously present, given
that their mismatch is simply $x+1/x$, the induced axion mass corresponds to
Eq.~(\ref{GenericmA}) with $n=1$, and is thus way too large at $m_{a^{0}}%
\sim\mathcal{O}(\Lambda_{\mathcal{B},\mathcal{L}}\times v/v_{s})$. This can be
understood qualitatively from the process depicted on the left in
Fig.~\ref{Fig1}, corresponding schematically to the symmetry-breaking terms
\begin{equation}
V_{Q_{1}\times Q_{2}}^{eff}=-m_{eff}^{2}\Phi_{2}^{\dagger}\Phi_{1}%
+h.c.\ ,\ m_{eff}^{2}\sim\frac{\Lambda_{reg}^{6}}{\Lambda_{\mathcal{B}%
,\mathcal{L}}^{4}}\times c_{1}^{IJKL}c_{2}^{ABKL}(\mathbf{Y}_{e}%
)^{AI}(\mathbf{Y}_{u})^{BJ}\ , \label{m12l5}%
\end{equation}
where $c_{i}$ are the Wilson coefficients of $Q_{i}$, and summation over the
flavor indices are understood. The scale $\Lambda_{reg}$ denotes that at which
the loop diagrams are regulated. In all UV scenarios we could think of, this
scale corresponds to that of the operators, $\Lambda_{reg}\approx
\Lambda_{\mathcal{B},\mathcal{L}}$. Indeed, if some new dynamics is introduced
that break the $U(1)_{1}\otimes U(1)_{2}$ symmetry, there is no reason not to
expect the same dynamics to induce corresponding breaking terms in the scalar sector.

If, instead of $Q_{2}$, one takes $Q_{1}$ together with $Q_{3}\equiv
e_{R}u_{R}^{2}d_{R}$, the mismatch in PQ charges is $2\times(x+1/x)$, and the
mass is $m_{a^{0}}\sim\mathcal{O}(v\times v/v_{s})$ from Eq.~(\ref{GenericmA})
with $n=2$. Again, this picture can be understood from the diagram on the
right in Fig.~\ref{Fig1}, with the corresponding dimension-four breaking term
in the effective potential:%
\begin{equation}
V_{Q_{1}\times Q_{3}}^{eff}=\frac{\lambda_{eff}}{2}(\Phi_{2}^{\dagger}\Phi
_{1})^{2}+h.c.\ \ ,\ \ \lambda_{eff}\sim\frac{\Lambda_{reg}^{4}}%
{\Lambda_{\mathcal{B},\mathcal{L}}^{4}}\times c_{1}^{IJKL}c_{3}^{ABCD}%
(\mathbf{Y}_{e})^{AI}(\mathbf{Y}_{u})^{BJ}(\mathbf{Y}_{u})^{CK}(\mathbf{Y}%
_{d})^{DL}\ .
\end{equation}
Thus, when $\Lambda_{reg}\approx\Lambda_{\mathcal{B},\mathcal{L}}$, the axion
mass becomes insensitive to the very high energy scale. Yet, it is still tuned
by the electroweak scale, and is thus far too large to solve the strong CP puzzle.

\paragraph{Scenario II: A viable scenario with many $\Delta\mathcal{B}$ and
$\Delta\mathcal{L}$ operators.}

The axion mass is too large for any combination of Weinberg operators carrying
different PQ charges. To get a viable scenario, we have to allow for operators
inducing different $\Delta\mathcal{B}$ and $\Delta\mathcal{L}$ patterns, so
that the effective potential term is forced to be of higher dimension. For
example, consider that instead of a dimension-six operator, $Q_{1}$ is
accompanied by the $\Delta\mathcal{B}=2$ dimension-nine operator $Q_{4}\equiv
d_{R}^{4}u_{R}^{2}$. Alone, $Q_{1}$ and $Q_{4}$ do not break the
$U(1)_{1}\otimes U(1)_{2}$ symmetry, since they have vanishing PQ charge for
some value of $\alpha$ and $\beta$. But if neutrinos have a Majorana mass
term, say $Q_{\nu}\equiv\ell_{L}^{2}\Phi_{i}^{2}$ of Eq.~(\ref{leffseesaw}),
then not all the $\Delta\mathcal{B}$ and $\Delta\mathcal{L}$ operators can be
simultaneously neutral. Thus, together, $Q_{1}$, $Q_{4}$, and $Q_{\nu}$
introduce too much $\Delta\mathcal{B}$ and $\Delta\mathcal{L}$ violation for
the axion to remain massless.

\begin{figure}[t]
\centering\includegraphics[width=0.90\textwidth]{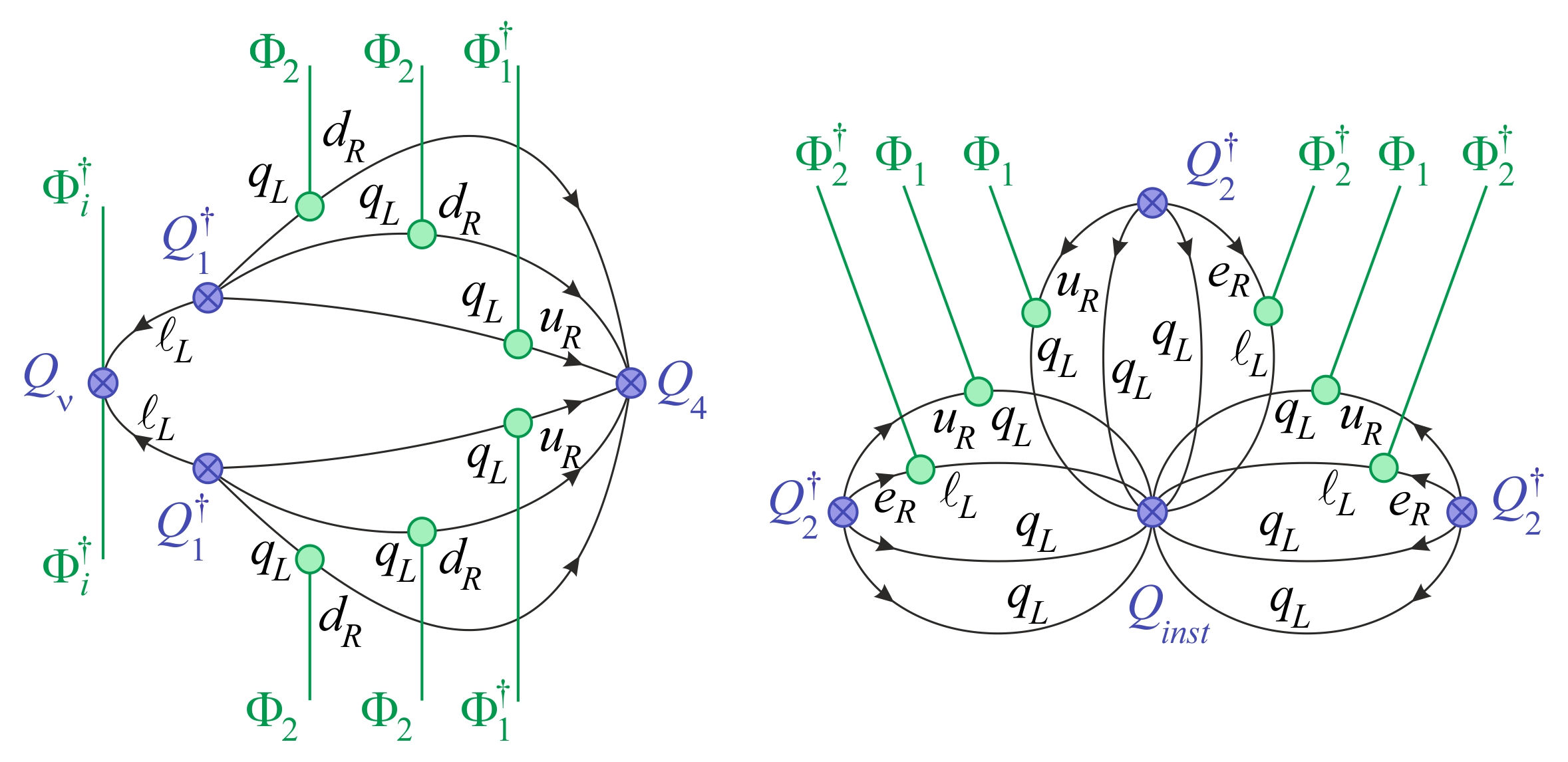}\caption{Left:
Fermion loop involving the dimension-five Majorana operator $Q_{\nu}=\ell
_{L}^{2}\Phi_{i}^{2}$, $i=1$ or $2$, the dimension-six Weinberg operator
$Q_{1}=\ell_{L}q_{L}^{3}$, and the $\Delta\mathcal{B}=2$ dimension-nine
operator $Q_{4}=d_{R}^{4}u_{R}^{2}$ . Right: Fermion loop involving the
instanton $Q_{inst}=q_{L}^{9}\ell_{L}^{3}$ interaction together with the
dimension-six $Q_{2}=e_{R}u_{R}q_{L}^{2}$ operator.}%
\label{Fig2}%
\end{figure}

Yet, the combined presence of these $\Delta\mathcal{B}$ and $\Delta
\mathcal{L}$ effects break $U(1)_{1}\otimes U(1)_{2}$ in a direction that can
be matched in the effective potential only at the cost of many Higgs doublets.
This combination of doublets need not be a power of $\Phi_{2}^{\dagger}%
\Phi_{1}$ anymore, and actually corresponds to the dimension-eight coupling
(see Fig.~\ref{Fig2})
\begin{equation}
V_{Q_{1}\times Q_{4}\times Q_{\nu}}^{eff}=\lambda_{eff}\ \Phi_{i}^{2}\Phi
_{1}^{2}\Phi_{2}^{\dagger4}+h.c.\ ,\ \ \lambda_{eff}\sim\frac{\Lambda
_{reg}^{6}}{\Lambda_{\mathcal{B},\mathcal{L}}^{10}}\times c_{\nu}c_{4}%
c_{1}^{\dagger}c_{1}^{\dagger}\mathbf{Y}_{u}^{2}\mathbf{Y}_{d}^{4}\ ,
\end{equation}
where $c_{\nu}$ is the Wilson coefficient of $Q_{\nu}$. Also, we have
suppressed flavor indices and identified the scale of all the operators to
$\Lambda_{\mathcal{B},\mathcal{L}}$ for simplicity. Clearly, the axion mass is
negligible in this case, of $\mathcal{O}(v^{3}/\Lambda_{\mathcal{B}%
,\mathcal{L}}^{2}\times v/v_{s})$ when $\Lambda_{reg}\approx\Lambda
_{\mathcal{B},\mathcal{L}}$. Even though $\theta_{eff}$ will not be entirely
disposed off, it is tiny and the strong CP puzzle is still solved.

Note that this estimate remain valid in the $\nu$DFSZ model even though
$Q_{\nu}$ is replaced by the singlet coupling to $\nu_{R}$. Indeed, the
leading term in the effective potential is then $\phi\Phi_{i}^{2}\Phi_{1}%
^{2}\Phi_{2}^{\dagger4}$, as can be seen from Fig.~\ref{Fig2} by splitting the
$\ell_{L}^{2}\Phi_{i}^{2}$ vertex into $\bar{\nu}_{R}\Phi_{i}\ell_{L}%
\otimes\phi\bar{\nu}_{R}^{C}\nu_{R}\otimes\bar{\nu}_{R}\Phi_{i}\ell_{L}$, and
one can check that this leads to the same estimate for the axion mass when
$v_{s}\sim\Lambda_{\mathcal{B},\mathcal{L}}$.

\paragraph{Scenario III: Electroweak instantons and the axion mass.}

As a final example, imagine now that only $Q_{2}$ and $Q_{\nu}$ are present.
At first sight, the $U(1)_{1}\otimes U(1)_{2}$ symmetry is preserved. However,
one still has to account for the electroweak instanton effects. Since
Eq.~(\ref{abInst}) is not compatible with the presence of $Q_{2}$, the axion
cannot be truly massless. It is a bit more tricky to estimate its mass in this
case because the electroweak instanton effects are not truly local. But to get
an idea of the induced mass, let us nevertheless use the same strategy as
above with $Q_{inst}=\ell_{L}^{3}q_{L}^{9}$. There will then be a new term in
the effective potential%
\begin{equation}
V_{Q_{2}\times Q_{inst}}^{eff}=\lambda_{eff}(\Phi_{2}^{\dagger}\Phi_{1}%
)^{3}+h.c.\ ,\ \ \lambda_{eff}\sim\frac{\Lambda_{reg}^{4}}{\Lambda
_{\mathcal{B},\mathcal{L}}^{6}}\times c_{inst}c_{2}^{\dagger3}(\mathbf{Y}%
_{e}\mathbf{Y}_{u}^{2}\mathbf{Y}_{d})^{3}\ .
\end{equation}
In this estimate, we consider that the UV regularization needs only to
compensate for the scale of the dimension-six operators, $\Lambda
_{\mathcal{B},\mathcal{L}}$, and not for the dimension-18 instanton effect.
So, this should be understood as nothing more than a rough estimate of the
maximal impact this combination of operators could have on the axion. In any
case, when $\Lambda_{reg}\approx\Lambda_{\mathcal{B},\mathcal{L}}$, the axion
mass is completely negligible because it is suppressed by the $\Lambda
_{\mathcal{B},\mathcal{L}}$ scale, see Eq.~(\ref{GenericmA}), because
instanton effects are tiny, $c_{inst}\sim\exp(-4\pi/g^{2})$, and because of
the flavor structure.\ Indeed, $Q_{inst}$ is fully antisymmetric in flavor
space, so first and second generation fermions circulate in the loop and there
will be many small Yukawa couplings. Actually, additional gauge interactions
may be needed to prevent the leading flavor contraction from vanishing, in a
way similar to what happens for the electroweak contribution to the EDMs, see
e.g. the discussion in Ref.~\cite{Smith:2017dtz}. Yet, even if tiny, this
shows that the axion would not be strictly massless in this case. Further, at
high temperature, when the QCD chiral symmetry is restored, these effects
would be dominant and force a specific alignment of the vacuum.

A very similar conclusion is encountered when non-perturbative quantum
gravity, which is expected to violate global symmetries, is taken into account
by adding non-local higher dimensional operators in the low energy effective
action~\cite{Barr:1992qq,Kamionkowski:1992mf}. Terms such as in
Eq.~(\ref{VBLeff}) are introduced with a Planck scale cut-off, $M_{P}%
\sim10^{19}$ GeV, implying a lower limit on their dimension $(2n+4)$ in order
not to impose permanently an alignment of the vacuum away from the strong CP solution.

\subsubsection{Dashen theorem approach}

The effective potential approach of the previous section is rather simple, but
it does not clearly show how the presence of too much $\mathcal{B}$ and/or
$\mathcal{L}$ violation imposes a realignment of the vacuum. This will be
described here, by perform the analysis directly in the broken phase.

Once the axion is introduced as the degree of freedom spanning the vacuum, the
fact that the symmetry is explicitly broken manifests itself via non-zero
matrix elements $\langle0|\mathcal{L}_{\mathcal{B},\mathcal{L}}|a^{0}\rangle$
and $\langle a^{0}|\mathcal{L}_{\mathcal{B},\mathcal{L}}|a^{0}\rangle$. The
latter corresponds to a mass term for the axion, and the former asks for a
realignment of the vacuum. Indeed, in the presence of the perturbation, the
vacuum is no lnger degenerate and the theory is unstable. It is only once at
the true vacuum, $|\Omega\rangle$, that the perturbation stops being able to
shift the vacuum and $\langle\Omega|\mathcal{L}_{\mathcal{B},\mathcal{L}%
}|a^{0}\rangle=0$. This condition on $|\Omega\rangle$ is equivalent to
Dashen's theorem~\cite{Dashen:1970et}, which states that the true vacuum is
that for which $\langle\Omega|\mathcal{L}_{\mathcal{B},\mathcal{L}}%
|\Omega\rangle$ is minimal.

Let us compute these matrix elements, and thereby the axion mass and true
vacuum $|\Omega\rangle$, for the specific case of the Weinberg operators,
$\mathcal{L}_{\mathcal{B},\mathcal{L}}=\mathcal{L}_{\mathcal{B},\mathcal{L}%
}^{\dim6}$.\ First, let us move to a more convenient basis. After the
reparametrization $\psi\rightarrow\exp(iPQ(\psi)a^{0}/v)\psi$ of the fermion
fields, the axion is removed from the Yukawa couplings. As detailed in
Ref.~\cite{Quevillon:2019zrd}, this generates derivative couplings
$-\partial_{\mu}a^{0}/v\times|J_{PQ}^{\mu}|_{fermions}$ from the fermion
kinetic terms, and anomalous non-derivative couplings $a^{0}/v\times
\partial_{\mu}J_{PQ}^{\mu}$ with $\partial_{\mu}J_{PQ}^{\mu}$ given in
Eq.~(\ref{dJPQTHDM}) from the non-invariance of the fermionic path integral
measure. In addition, since $\mathcal{L}_{\mathcal{B},\mathcal{L}}$ is not
invariant under $U(1)_{PQ}$, each operator gets transformed into
$Q_{i}\rightarrow Q_{i}\exp(iPQ(Q_{i})a^{0}/v)$.

Forgetting for now the anomalous couplings, the only couplings surviving in
the static limit are the non-derivative couplings with the $\Delta
(\mathcal{B}+\mathcal{L})$ interactions,%
\begin{align}
\mathcal{L}_{\mathcal{B},\mathcal{L}}^{\dim6}  &  =\frac{c_{1}}{\Lambda^{2}%
}\ell_{L}q_{L}^{3}\exp\left(  i(3\alpha+\beta)\frac{a^{0}}{v}\right)
+\frac{c_{2}}{\Lambda^{2}}e_{R}u_{R}^{2}d_{R}\exp\left(  i\left(
3\alpha+\beta+2x+\frac{2}{x}\right)  \frac{a^{0}}{v}\right) \nonumber\\
&  +\left(  \frac{c_{3}}{\Lambda^{2}}e_{R}u_{R}q_{L}^{2}+\frac{c_{4}}%
{\Lambda^{2}}\ell_{L}q_{L}d_{R}u_{R}\right)  \exp\left(  i\left(
3\alpha+\beta+x+\frac{1}{x}\right)  \frac{a^{0}}{v}\right)  +h.c.\ .
\end{align}
When taken alone, none of these operators is able to induce $\langle
\Omega|\mathcal{L}_{\mathcal{B},\mathcal{L}}^{\dim6}|a^{0}\rangle$ or $\langle
a^{0}|\mathcal{L}_{\mathcal{B},\mathcal{L}}^{\dim6}|a^{0}\rangle$. For
example, with only $Q_{1}$, the simplest $\Delta(\mathcal{B}+\mathcal{L})=0$
matrix element arises from a $Q_{1}^{\dagger}\otimes Q_{1}$ combination, and
the axion field disappears. Some interference between two or more operators
with different phases is needed. Let us consider that arising from $Q_{1}$ and
$Q_{2}$. We have two contributions, $Q_{1}^{\dagger}\otimes Q_{2}$ and
$Q_{2}^{\dagger}\otimes Q_{1}$. Since we are only after $\Delta(\mathcal{B}%
+\mathcal{L})=0$ matrix elements with external axion fields, we can consider
the generating function%
\begin{align}
\mathcal{V}_{\mathcal{B},\mathcal{L}}^{\dim6}  &  =\langle\Omega|Q_{1}\otimes
Q_{2}^{\dagger}|\Omega\rangle\exp\left(  -2i\left(  x+\frac{1}{x}\right)
\frac{a^{0}}{v}\right)  +h.c.\nonumber\\
&  =2|\langle\Omega|Q_{1}\otimes Q_{2}^{\dagger}|\Omega\rangle|\cos\left(
\delta_{12}+2\left(  x+\frac{1}{x}\right)  \frac{a^{0}}{v}\right) \nonumber\\
&  =2|\langle0|Q_{1}\otimes Q_{2}^{\dagger}|0\rangle|\cos\left(  \delta
_{12}+2\left(  x+\frac{1}{x}\right)  \frac{a^{0}+\omega}{v}\right)  \ ,
\end{align}
where $\delta_{12}$ denotes the phase of $\langle\Omega|Q_{1}\otimes
Q_{2}^{\dagger}|\Omega\rangle$. In the last line, we use the fact that the
vacuum space is spanned by the axion, i.e., any two vacua are related by
shifts in the axion field. This permits to trade $|\Omega\rangle$ for the free
parameter $\omega$.

Expanding the cosine function up to second order, the axion mass is found to
be consistent with the previous estimate, Eq.~(\ref{m12l5}), since
$\langle0|Q_{1}\otimes Q_{2}^{\dagger}|0\rangle$ corresponds to the diagrams
of Fig.~\ref{Fig1} with the external Higgs fields replaced by their vacuum
expectation values. Concerning the vacuum, $\langle\Omega|\mathcal{L}%
_{\mathcal{B},\mathcal{L}}^{\dim6}|a^{0}\rangle$ is obtained from
$\partial\mathcal{V}_{\mathcal{B},\mathcal{L}}^{\dim6}/\partial a^{0}$ at
$a^{0}=0$, and thus vanishes when $\omega$ satisfies%
\begin{equation}
\delta_{12}+2\left(  x+\frac{1}{x}\right)  \frac{\omega}{v}=0. \label{AlignBL}%
\end{equation}
The fact that the prefered direction is set by the phase of $\langle
\Omega|Q_{1}\otimes Q_{2}^{\dagger}|\Omega\rangle$ can be understood as
follow. In the absence of $\mathcal{L}_{\mathcal{B},\mathcal{L}}^{\dim6}$,
thanks to the still exact $U(1)_{PQ}$ symmetry, one can remove any phase
occurring in the fermion mass terms as well as take real VEVs, $v_{1,2}$ for
the two Higgs doublets (see Eq.~(\ref{PolPara})). But, it is no longer
possible to keep both VEVs real once $U(1)_{PQ}$ is broken by $\mathcal{L}%
_{\mathcal{B},\mathcal{L}}^{\dim6}$, and the specific choice in
Eq.~(\ref{AlignBL}) becomes compulsory.

In some sense, we can also understand $\mathcal{V}_{\mathcal{B},\mathcal{L}%
}^{\dim6}$ as a contribution to the effective potential of the axion. With
this picture, bringing back the anomalous couplings and turning on the QCD
effects, the full axion potential looks like $\mathcal{V}_{eff}=\mathcal{V}%
_{\mathcal{B},\mathcal{L}}^{\dim6}+\mathcal{V}_{QCD}$ with $\mathcal{V}%
_{QCD}\sim m_{\pi}^{2}f_{\pi}^{2}\bar{m}\cos(\theta_{QCD}+a^{0}/v)$ and
$\bar{m}=m_{u}m_{d}/(m_{u}+m_{d})^{2}$. Thus, the strong CP puzzle is solved
only if $\mathcal{V}_{QCD}$ dominates and forces the vacuum to align itself to
kill $\theta_{QCD}$. In the present case, given that the $\mathcal{V}%
_{\mathcal{B},\mathcal{L}}^{\dim6}$-induced axion mass is much larger than
that induced by $\mathcal{V}_{QCD}$, the constraint from $\mathcal{V}%
_{\mathcal{B},\mathcal{L}}^{\dim6}$ is stronger and the vacuum is rather
aligned in the direction of Eq.~(\ref{AlignBL}), leaving the strong CP-puzzle open.

\section{Conclusions\label{Ccl}}

Axion models are based on the spontaneous breaking of an extra $U(1)$
symmetry. When this symmetry has a strong anomaly, the associated Goldstone
boson, the axion, ends up coupled to gluons, and this ensures the strong CP
violation relaxes to zero in the non-perturbative regime. In this paper, we
analyzed more specifically the PQ and DFSZ axion models, where SM fermions as
well as the Higgs fields responsible for the electroweak symmetry breaking are
charged under the additional $U(1)$ symmetry.\ A characteristic feature of
these models is that the true $U(1)_{PQ}$ symmetry corresponding to the axion
is not trivial to identify, because of the presence of three other $U(1)$
symmetries acting on the same fields: baryon number $\mathcal{B}$, lepton
number $\mathcal{L}$, and weak hypercharge. As a consequence, in general, the
PQ charges can be defined only after $U(1)_{Y}$ is spontaneously broken, and
even then, those of the fermions remain ambiguous whenever baryon or lepton
number is conserved. Our purpose was to study this ambiguity, see when it can
be lifted, and how it leaves the axion phenomenology intact. Our main results are:

\begin{itemize}
\item The ambiguities in the PQ charges of the fermions, here parametrized by
$\alpha$ and $\beta$, is well-known but it is often interpreted as a freedom.
One seems free to fix $\alpha$ and $\beta$ as one wishes. Doing this, however,
prevents any further analysis of $\mathcal{B}$ and $\mathcal{L}$ violation.
For example, if one chooses to assign PQ charges only to right-handed
fermions, a $\Delta\mathcal{B}=2$ operator like $(u_{R}d_{R}d_{R})^{2}$ would
be forbidden. Yet, this is merely a consequence of the choice made for the PQ
charges. What we showed here is that it is compulsory to keep the fermion
charge ambiguity explicit to leave the theory the necessary room to adapt to
the presence of $\mathcal{B}$ and/or $\mathcal{L}$ violation. Indeed, in the
presence of such interactions, these ambiguities automatically disappear, and
the corresponding parameters $\alpha$ and $\beta$ are fixed to specific
values, when the $U(1)_{PQ}$ symmetry aligns itself with the remaining $U(1)$
symmetry of the Lagrangian. This proves that such violations of $\mathcal{B}$
and $\mathcal{L}$ can be compatible with the PQ symmetry.

\item Since there are two parameters, reflecting the two accidental symmetries
$U(1)_{\mathcal{B}}\otimes U(1)_{\mathcal{L}}$, axion models can accommodate
for breaking terms in two independent directions. This means for example that
adding a $\Delta\mathcal{L}=2$ Majorana mass for the neutrinos as well as some
$\mathcal{B}+\mathcal{L}$ violating operators, say $e_{R}u_{R}q_{L}^{2}$ and
$\ell_{L}q_{L}d_{R}u_{R}$, preserves the axion solution to the strong CP
puzzle. Yet, this compatibility is delicate and needs to be checked in
details. For example, adding both the operators $\ell_{L}q_{L}^{3}$ and
$e_{R}u_{R}q_{L}^{2}$ spoils the axion solution completely, even though these
operators have the same $\mathcal{B}$ and $\mathcal{L}$ quantum numbers. When
they are both present, there is simply not enough room for the $U(1)_{PQ}$
symmetry to remain active.

\item In many cases, the capability of axion models to accommodate for
$\mathcal{B}$ and $\mathcal{L}$ violation is saturated from the start. Indeed,
first, these models should be compatible with neutrino masses, and seesaw
mechanisms being the most natural, some $\Delta\mathcal{L}=2$ effects are
present. Second, electroweak instantons generate $\mathcal{B}+\mathcal{L}$
violating effects, and even if negligible at low energy, their mere existence
forces the PQ symmetry to be realized in a specific way. In this case, there
remains not much room for other $\mathcal{B}$ and/or $\mathcal{L}$ violating
effects. For example, the axion cannot remain a true Goldstone boson in the
presence of say the $e_{R}u_{R}q_{L}^{2}$ or $(u_{R}d_{R}d_{R})^{2}$ operator.
Yet, the instanton interaction is so small that the induced mass of the axion
is well below the QCD mass, and the strong CP puzzle is still solved.
Obviously, the situation changes at high temperature. If the electroweak
contribution to the axion mass becomes larger than the QCD contribution, the
axion is initially not aligned in the CP-conserving direction but does so only
at a later time. Such a situation could have important cosmological consequences.

\item Usually, axion models are specified in a particular representation, in
which the axion has only derivative couplings to SM fermions, and anomalous
couplings to gauge field strengths. Because these effective couplings arise
from chiral rotations of the fermion fields, tuned by their PQ charges, some
dependences on $\alpha$ and $\beta$ are introduced (explicitly or implicitly)
in the Lagrangian. At the same time, we have shown that $\alpha$ and $\beta$
take on various very different values, depending on the $\Delta\mathcal{B}$
and/or $\Delta\mathcal{L}$ effects present. So, the axion effective
interactions are strongly dependent on the presence of these $\Delta
\mathcal{B}$ and/or $\Delta\mathcal{L}$ interactions, whatever their intrinsic
size. In this respect, the electroweak couplings $a^{0}W_{\mu\nu}^{i}\tilde
{W}^{i,\mu\nu}$, $i=1,2,3$, is extreme in that the theory turns it off
automatically whenever the PQ current has to circumvent the tiny electroweak
instanton interactions. Of course, these dependences on $\alpha$ and $\beta$
are spurious. As we demonstrated in Ref.~\cite{Quevillon:2019zrd}, the
$\alpha$ and $\beta$ terms occurring in the derivative interactions always
cancel out exactly with those of the anomalous interactions, and the physical
axion to fermion or gauge boson amplitudes are independent of $\alpha$ and
$\beta$. In particular, the $a^{0}W^{+}W^{-}$ coupling is non-zero even when
the anomalous $a^{0}W_{\mu\nu}^{i}\tilde{W}^{i;\mu\nu}$ term is forced out of
the axion effective Lagrangian by electroweak instantons.

\item Several scenarios were discussed: the PQ and DFSZ axion with massless
neutrinos, with a seesaw mechanism of type I and of type II, and the $\nu$DFSZ
where the singlet also plays the role of the majoron. Then, additional
requirements were discussed, arising from the electroweak instantons, a GUT
constraint, or various $\mathcal{B}$ violating operators. Despite their
variety, for all those settings, the PQ charges of the two Higgs doublets and
the fermions are the same, up to specific values for $\alpha$ and $\beta$, and
up to negligible corrections in the type II seesaw. Though this can be
understood as the orthogonality condition among Goldstone bosons stays
essentially the same and the Yukawa couplings are always those of
Eq.~(\ref{YukQuark}), it is often obscured by the normalization of the PQ
charges. Yet, this is remarkable because it means the low-energy phenomenology
of the axion is the same in all these models, since it is independent of
$\alpha$ and $\beta$.\ This is most evident adopting a linear parametrization
for the two Higgs doublets, since the axion then does not couple directly to
gauge bosons, while its coupling to each fermion is simply proportional to the
fermion mass times the PQ charge of the doublet to which it
couples~\cite{Quevillon:2019zrd}.
\end{itemize}

The results of this paper should have implications in other settings where
$\mathcal{B}$ and/or $\mathcal{L}$ violations occur, most notably in
supersymmetry if R-parity is not conserved and in Grand Unified Theories.
While embedding the axions in those models has already been proposed, further
work to identify the most promising scenario is required~\cite{toappear}. In
this respect, the connection with cosmology, either via the axion relic
density or its possible impact on baryogenesis, could provide invaluable information.

\end{document}